\documentclass[twocolumn, pra, superscriptaddress,notitlepage]{revtex4-1}
\usepackage{graphicx}
\usepackage{physics}
\usepackage{epsfig}
\usepackage{color}
\usepackage{xspace}
\usepackage{array}
\usepackage{amsmath}
\usepackage{amsfonts}
\usepackage{ulem}
\usepackage[dvipsnames]{xcolor}
\usepackage{colortbl}
\usepackage[breaklinks,colorlinks,linkcolor=blue,citecolor=blue,urlcolor=blue]{hyperref}

\begin{document}
\title{Scaling Relations of Spectrum Form Factor and Krylov Complexity at Finite Temperature}

 \author{Chengming Tan}
 \affiliation{Hefei National Research Center for Physical Sciences at the Microscale and School of Physical Sciences,University of Science and Technology of China, Hefei 230026, China}

 \author{Zhiyang Wei}
\affiliation{MOE Key Laboratory for Nonequilibrium Synthesis and Modulation of Condensed Matter, Shaanxi Province Key Laboratory of Quantum
Information and Quantum Optoelectronic Devices, School of Physics, Xi’an Jiaotong University, Xi’an 710049, China}

 \author{Ren Zhang}
 \email{ renzhang@xjtu.edu.cn}
\affiliation{MOE Key Laboratory for Nonequilibrium Synthesis and Modulation of Condensed Matter, Shaanxi Province Key Laboratory of Quantum
Information and Quantum Optoelectronic Devices, School of Physics, Xi’an Jiaotong University, Xi’an 710049, China}
\affiliation{Hefei National Laboratory, Hefei, 230088, China}
\date{\today}

\begin{abstract}
In the study of quantum chaos diagnostics, considerable  attention has been attributed to the Krylov complexity and spectrum form factor (SFF) for systems at infinite temperature. These investigations have unveiled universal properties of quantum chaotic systems. By extending the analysis to include the finite temperature effects on the Krylov complexity and SFF, we demonstrate that the Lanczos coefficients $b_n$, which are associated with the Wightman inner product, display consistency with the universal hypothesis presented in PRX 9, 041017 (2019). This result contrasts with the behavior of Lanczos coefficients associated with the standard inner product. Our results indicate that the slope $\alpha$ of the $b_n$ is bounded by $\pi k_BT$, where $k_B$ is the Boltzmann constant and $T$ the temperature.
We also investigate the SFF, which characterizes the two-point correlation of the spectrum and encapsulates an indicator of ergodicity denoted by $g$ in chaotic systems. Our analysis demonstrates that as the temperature decreases, the value of $g$ decreases as well. Considering that $\alpha$ also represents the operator growth rate, we establish a quantitative relationship between ergodicity indicator and Lanczos coefficients slope.
To support our findings, we provide evidence using the Gaussian orthogonal ensemble and a random spin model. Our work deepens the understanding of the finite temperature effects on Krylov complexity, SFF, and the connection between ergodicity and operator growth.
\end{abstract}

\maketitle
\section{introduction}
Unveiling the scaling relations of physical quantities is an important and fundamental topic in the study of classical and quantum physics~\cite{SR5,SR1,SR2,SR3,SR4,MBLS1,PT1}. In the context of equilibrium systems, significant progress has been made in this direction, leading to a deeper understanding of critical phenomena, phase transitions, and universal behaviors~\cite{SRnew1,SRnew2,SRnew3,SRnew4}. The investigations of scaling relations in numerous systems are still attracting growing interests, especially in the context of non-equilibrium systems and quantum dynamics of chaotic systems~\cite{ergo1,ergo2}.

Quantum chaos has been an attractive topic across diverse disciplines, spanning from condensed matter physics to black hole physics and high-energy physics~\cite{BH-SYK1,BH-SYK2,chaos_2017}. This field holds the potential to offer deeper insights into phenomena like thermalization~\cite{ETH1,ETH2,ETH3,ETH4,ETH5,ETH6}, information scrambling~\cite{IS1,IS2,IS3,IS4}, and quantum entanglement~\cite{ETH6,QEntangle1,QEntangle2}. However, comprehensively diagnosing quantum chaos remains a formidable challenge. To date, several methods have emerged for characterizing the nature of quantum chaos. These include spectrum statistics~\cite{ES2,ES1,ES3}, out-of-time-correlation~\cite{OTOC1,OTOC3}, Loschmidt echo~\cite{IS1}, and complexity~\cite{OC1}. Each diagnostic tool captures different aspects of quantum chaos and provides valuable insights individually. Recently, spectrum form factor (SFF)~\cite{BH-SYK1,chaos_2017,DRP1,SFF1,SFF2} and Krylov complexity (K-complexity) ~\cite{Ehud} stand out as prominent examples, offering perceptions into the energy spectrum and operator growth in quantum chaotic systems.

Typically, the characteristics of quantum chaos manifest across the entire spectrum. To quantify the chaos strength, the statistical properties are conventionally studied.  For example, the Wigner-Dyson distribution describes the statistical properties of a quantum chaotic system at infinite temperature, where a level repulsion presents in the energy spectrum~\cite{WD1,RM,QCS2010}. The statistical properties of the spectrum provide information solely about the nearest level spacing within the spectrum. In contrast, the SFF measures energy correlations between any two states, which allows it to go beyond the constraints of spectrum statistics and capture richer information. SFF has found widespread use in diverse contexts, including random matrix theory (RMT)~\cite{WD1,RM1,RM,QCS2010,RM2}, the Sachdev-Ye-Kitaev model~\cite{chaos_2017,BH-SYK1}
, and the Hubbard model~\cite{Hubbard,Hubbard2,Hubbard1}. The intriguing dip-ramp-plateau structure observed in many chaotic systems indicats the universality of SFF~\cite{DRP1,DRP2,DRP3}. The SFF has also been generalized to encapsulate the correlation between multi energy levels~\cite{ES3,GSFF1,GSFF2,GSFF3}.

 In quantum systems, the chaotic feature is also manifested through operator growth. Specifically, the evolution of an initially simple operator $\hat{\cal O}(0)$ gradually becomes more complex under the Heisenberg evolution. This is because the operator $\hat{\cal O}(0)$ does not commute with the Hamiltonian $\hat{H}$ of an interacting quantum system. Several approaches have been proposed to quantify the operator growth~\cite{OG2,OG4,OG6,operator-size}.
Recently, the concept of K-complexity was introduced to characterize operator growth~\cite{Ehud} and has been applied to various systems, including strongly correlated systems, quantum circuits, and black hole physics~\cite{Krylov6,Krylov7,Krylov12,Hui_Krylov,KC_OC1,KC_OC2,KC_OC3,KC_OC4,Krylov1,Krylov2,Krylov3,Krylov4,Krylov9,Krylov11,Krylov13,Krylov14,Krylov15,Krylov16,Krylov18,Krylov19,Krylov20,Krylov21,Krylov22,Krylov23,Krylov24,Krylov25,Krylov26,Hubbard3,KC_MBS,Krylov8,Krylov10,Krylov17}. It was found that operator growth can be quantified by the mean value of wave packet position on a half-infinite chain. Motivated by this idea, a universal hypothesis of operator growth has also been proposed, where the Lanczos coefficients can be used to diagnose the presence of quantum chaos. Meanwhile, the autocorrelation function ${\cal C}(t)={\rm Tr}(\hat{\cal O}(t)\hat{\cal O}(0))$ defined at infinite temperature can be readily obtained and has been intensively studied in various contexts~\cite{autoC3,autoC1,autoC2}. It is surprising to find that the universal behavior of the autocorrelation function is determined by the first few Lanczos coefficients~\cite{UHAF-z}. In addition, some studies have explored the relations between K-complexity and other complexity measures, such as circuit complexity, leading to a geometric interpretation of K-complexity~\cite{Geometry_Krylov1,CC-KC,Geometry_Krylov2}. Furthermore, the concept of K-complexity is extended to systems at finite temperatures, and relevant discussions are presented \cite{Ehud,Krylov8,Krylov10,Krylov17,autoC3,Temperature1,Temperature3,Temperature2}.

In this work, we establish a connection between K-complexity and SFF at finite temperature. Our main conclusions can be summarized as follows: (1) The use of the Wightman inner product and the standard inner product results in distinct Lanczos coefficients $b_{n}$, and the slope $\alpha$ of $b_{n}$ in the former is limited by $\pi k_BT$; (2) The ergodicity indicator $g$ diminishes with decreasing temperature; (3) We establish a quantitative relationship between $g$ and $\alpha$ after proper scaling.

\section{Lanczos coefficients at finite temperature}
Let us begin with a brief review of the derivation of Lanczos coefficients, and highlight the finite temperature effect of operator growth. Consider a scenario where we have a simple operator $\hat{\cal O}$ and a Hamiltonian $\hat{H}$. Typically, in the presence of interactions, they do not commute. The evolution of the operator $\hat{\cal O}$ follows the Heisenberg equation $-i\partial_{t}\hat{\cal O}(t)=[\hat{H},\hat{\cal O}(t)]$, and the formal solution reads $\hat{\cal O}(t)={\rm e}^{i\hat{H}t}\hat{\cal O}{\rm e}^{-i\hat{H}t}$. Henceforth, we set $\hbar=1$. Therefore, the initial simple operator becomes progressively more complex due to the noncommutative relation between $\hat{\cal O}(t)$ and $\hat{H}$, which is also termed as operator growth.
In order to quantify the complexity of $\hat{\cal O}(t)$, the concept of K-complexity was introduced by Parker, {\it et.al.}~\cite{Ehud}. Given a Hilbert space spanned by ${\ket{i}}$, the operator $\hat{\cal O}=\sum_{i,j}{\cal O}_{ij}\ket{i}\bra{j}$ can be mapped to an ``operator state'' $|{\hat{\cal O}}\rangle=\sum_{i,j}{\cal O}_{ij}\ket{i}\otimes\ket{j}$ living in the double Hilbert space spanned by ${\ket{i}\otimes\ket{j}}$. By defining a Liouvillian superoperator $\hat{\cal L}=[\hat{H},\cdots]$, the Heisenberg equation can be rewritten as $\partial_{t}|{\hat{\cal O}(t)}\rangle=i\hat{\cal L}|{\hat{\cal O}(t)}\rangle$.
In the double Hilbert space, the Liouvillian superoperator appears as a conventional operator, and is written as $\hat{\cal L}=\hat{H}\otimes\hat{\mathbb I}-\hat{\mathbb I}\otimes\hat{H}^{T}$ where $\hat{\mathbb I}$ denotes the identity operator with the same dimension of $\hat{H}$. The superscript $T$ denotes transpose. The time-dependent operator state is expressed as:
\begin{align}
\label{Ot}
|\hat{\cal O}(t)\rangle={\rm e}^{i\hat{\cal L}t}|{\hat{\cal O}}\rangle=\sum_{n=0}^{\infty}\frac{(it)^{n}}{n!}|{\hat{\cal L}^{n}\hat{\cal O}}\rangle.
\end{align}
This means that $|{\hat{\cal O}(t)}\rangle$ is expanded under the basis spanned by ${|{\hat{\cal L}^{n}\hat{\cal O}}}\rangle$. It is important to note that this basis is neither normalized nor orthogonal. To address this issue, the Krylov basis is defined by the Gram-Schmidt orthogonalization. It is pointed out that one must properly define the inner product of two operator states as follows:
\begin{align}
\langle{\hat A}|{\hat B}\rangle_{\beta}^{g}=\frac{1}{Z}\int_{0}^{\infty}g(\lambda){\rm Tr}({\rm e}^{-\beta \hat H} {\rm e}^{\lambda \hat H} \hat A^{\dagger}{\rm e}^{-\lambda\hat H}\hat B)d\lambda.
\end{align}
Here, $Z={\rm Tr}({\rm e}^{-\beta\hat H})$ denotes the partition function, $\beta=1/(k_{\rm B}T)$ represents the inverse temperature, and $k_{\rm B}$ indicates the Boltzmann constant. $g(\lambda)$ satisfies the conditions $g(\lambda)\geq0$, $g(\beta-\lambda)=g(\lambda)$, and $\beta^{-1}\int_{0}^{\beta}d\lambda g(\lambda)=1$ ~\cite{Ehud}. Different choices of $g(\lambda)$ define distinct inner products. In this work, we consider two specific choices: (1) The standard inner product is defined by choosing $g(\lambda)=[\delta(\lambda)+\delta(\lambda-\beta)]/2$, and is explicitly written as,
\begin{align}
\langle{\hat A}|{\hat B}\rangle_{\beta}^{S}=\frac{1}{2Z}{\rm Tr}\left({\rm e}^{-\beta\hat H}\hat A^{\dagger}\hat B+ \hat A^{\dagger}{\rm e}^{-\beta\hat H}\hat B\right).
\end{align}
Here, by ``standard'' we mean that this kind of inner product is widely adopted in the linear response theory.
(2) Another inner product often considered in quantum field theory is the Wightman inner product, defined by $g(\lambda)=\delta(\lambda-\beta/2)$, and is explicitly written as,
\begin{align}
\langle{\hat A}|{\hat B}\rangle_{\beta}^{W}=\frac{1}{Z}{\rm Tr}\left(e^{-\beta\hat H/2}\hat A^{\dagger}e^{\beta\hat H/2}\hat B\right).
\end{align}
It should be noted that at infinitely high temperature, i.e., $\beta\to0$, both the standard and Wightman inner products reduce to the same one: $\langle{\hat A}|{\hat B}\rangle={\rm Tr}\left(\hat A^{\dagger}\hat B\right)/{\rm Tr}(\hat {\mathbb I})$.
With the inner product, we define the Krylov basis ${|{\hat{\cal O}_{n}}}\rangle$ using the Lanczos algorithm as follows: for $n=1,2$, we have $|{\hat{\cal O}_{0}}\rangle=|{\hat{\cal O}}\rangle$ and $|{\hat{\cal O}_{1}}\rangle=b_{1}^{-1}\hat{\cal L}|{\hat{\cal O}_{0}}\rangle$, respectively, where $b_{1}=\langle{\hat{\cal L}\hat{\cal O}_{0}}|{\hat{\cal L}\hat{\cal O}_{0}}\rangle^{1/2}$; then for $n>2$, $|{\hat{\cal O}_{n}}\rangle$ is defined inductively as,
\begin{align}
|{\hat{\cal A}_{n}}\rangle=&\hat{\cal L}|{\hat{\cal O}_{n-1}}\rangle-b_{n-1}|{\hat{\cal O}_{n-2}}\rangle;\nonumber\\
b_{n}=&\left[\langle{\hat {\cal A}_{n}}|{\hat {\cal A}_{n}}\rangle_{\beta}^{g}\right]^{1/2};\\
|{\hat{\cal O}_{n}}\rangle=&b_{n}^{-1}|{\hat{\cal A}_{n}}\rangle.\nonumber
\end{align}
Here, $b_{n}$'s indicate the Lanczos coefficients, which obviously depend on temperature. Previous research has demonstrated that the asymptotic behavior of these coefficients at large $n$ serves as a diagnostic for chaotic systems~\cite{Ehud}. Using the completeness and orthogonality of the Krylov basis, the Heisenberg equation maps to a discrete Schr\"odinger-like equation $\partial_{t}\varphi_{n}(t)=b_{n}\varphi_{n-1}(t)-b_{n+1}\varphi_{n+1}(t)$ with $\varphi_{n}(t)=i^{-n}\langle{\cal O}_{n}|{\cal O}(t)\rangle$. The natural initial condition reads $\varphi_{n}(0)=\delta_{n0}$.

\begin{figure}[t]
    \centering
    \includegraphics[width=0.48\textwidth]{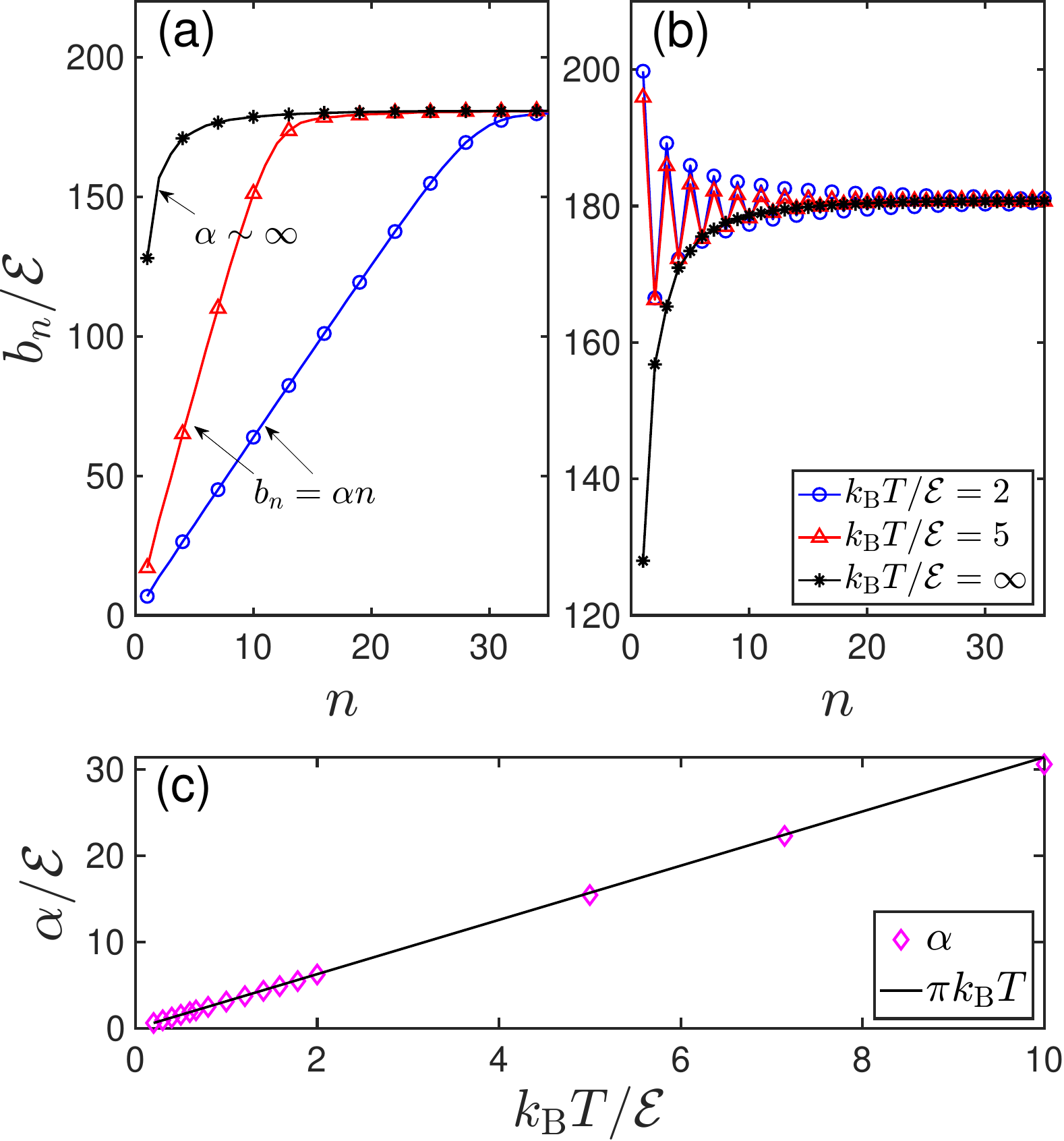}
    \caption{Finite temperature Lanczos coefficients $b_{n}$ for Gaussian orthogonal ensemble (GOE). The operator is chosen as $\hat{\sigma}_{z}\otimes\hat{\mathbb I}_{L/2}$, and the dimension of random matrix is $8192$. In (a) and (b), Wightman and standard inner product are used, respectively. (a):  At finite temperature, $\{b_{n}\}$ defined by Wightman inner product shows linear behavior, which is consistent with the universal hypothesis. (b): The finite temperature $b_{n}$ presents oscillatory behavior which is in contrast to the universal hypothesis. (c): The slope $\alpha(T)=\pi k_{\rm B}T$ aligns with upper bound of the operator growth rate, and approach at infinite temperature.}
     \label{finiteTbn}
\end{figure}

To illustrate the difference in ${b_{n}}$'s defined by the standard and Wightman inner products, we calculate ${b_{n}}$ for the Gaussian orthogonal ensemble (GOE), which is an ideal chaotic model. The Hamiltonian is expressed as $\hat{H}_{\rm GOE} = {\rm GOE}(L,\mu,\sigma)$, where ${\rm GOE}(L,\mu,\sigma)$ represents a $L\times L$ Hermitian matrix of which the elements are sampled from Gaussian distribution with mean value $\mu$ and standard deviation $\sigma$. In our case, we choose $\mu = 0$ and $\sigma=\mathcal{E}$, where the $\mathcal{E}$ is the characteristic energy.
The operator is chosen as  $\hat{\sigma}_{z}\otimes\hat{\mathbb I}_{L/2}$, where $\hat{\sigma}_{z}$ and $\hat{\mathbb I}_{L/2}$ denote the Pauli matrix and $L/2\times L/2$ identity matrix, respectively.  Obviously, this is a simple operator. By simple, we mean it is represented by a sparse matrix. In evolution, this operator becomes complex progressively. The ${b_{n}}$'s obtained using the Wightman and standard inner products are shown in Fig.~\ref{finiteTbn}(a) and (b), respectively.   It is important to note that at infinite temperature, both inner products yield the same ${b_{n}}$ values, aligning with our expectations. However, the $b_{n}$'s at finite temperature showcase quite different behaviors for the above-mentioned inner products. The Wightman inner product leads to a linear $b_{n}$ at small $n$, i.e., $b_{n}\approx \alpha(T) n$. The slope $\alpha$ increases as the temperature rises and approaches infinity at infinite temperature. This is reasonable in the sense that at higher temperatures, the speed of operator growth is larger. The range of $n$ that supports linear $b_{n}$ also extends at low temperatures. This indicates that the universal hypothesis is applicable even at finite temperature when the Wightman inner product is adopted. The saturation at large $n$ is due to finite-size effects. In contrast, $b_{n}$ shows oscillatory behavior at finite temperature when the standard inner product is used, and the oscillatory behavior becomes more pronounced at lower temperatures, which is not aligned with the universal hypothesis. In the following analysis, we will focus on the Wightman inner product. In Fig.~\ref{finiteTbn}(c), we present the slope $\alpha(T)$ for various temperatures. It is clear that for GOE,
\begin{align}
\alpha(T)=\pi k_BT,
\label{Eq,alpha-T}
\end{align}
which, on the one hand, is consistent with the universal hypothesis on the upper bound of the Lanczos coefficient growth rate $\alpha(T)\leq \pi k_BT$, and on the other hand, manifests GOE being the most chaotic model. The small derivation at high temperature is due to the finite size effect.

\section{Spectrum form factor}
\begin{figure}[t]
	\centering	
	\includegraphics[width=0.48\textwidth]{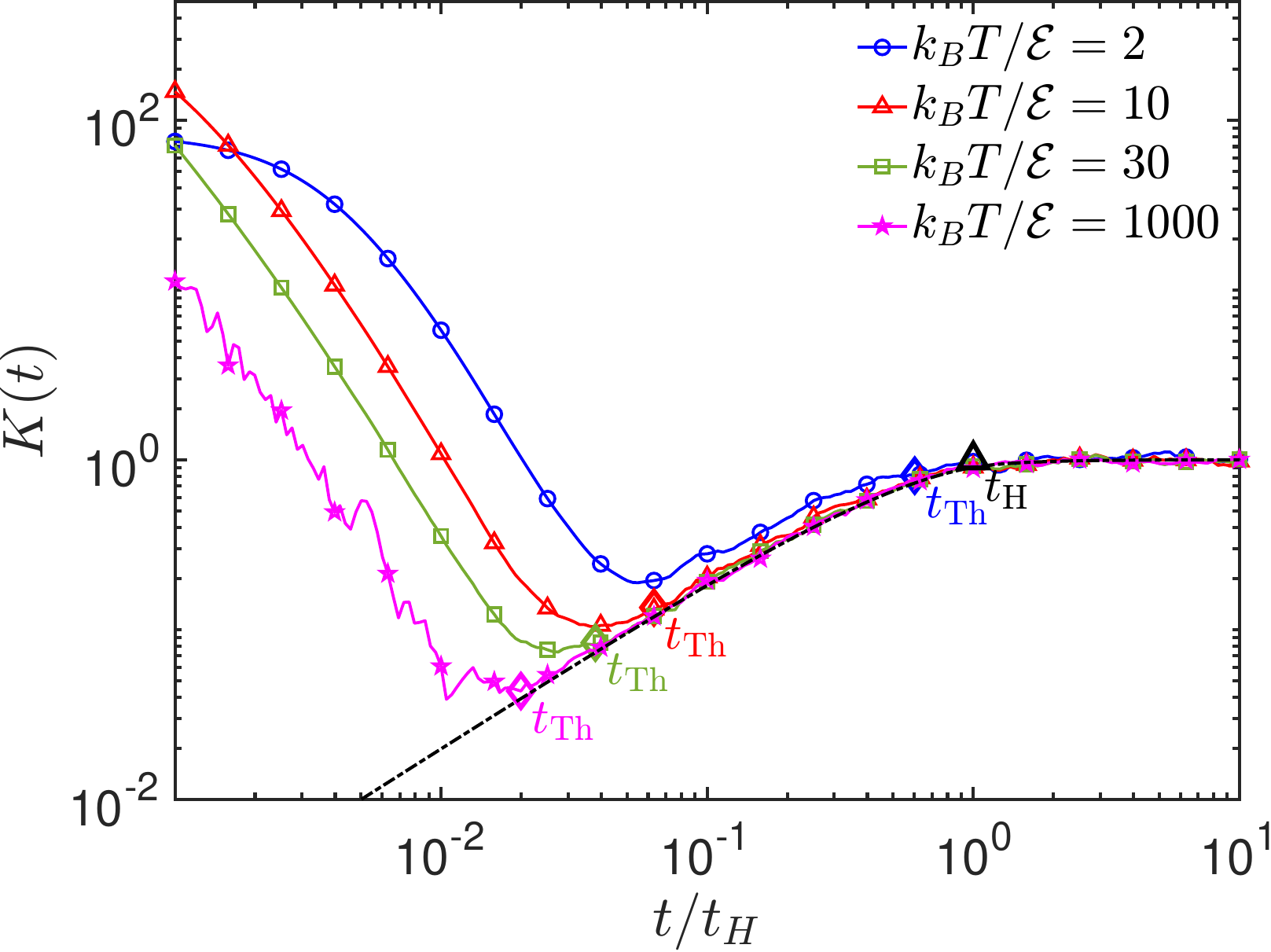}
	\caption{The spectrum form factor(SFF) at finite temperature.  The ``dip-ramp-plateau'' structure of SFF survives at finite temperature(colored line). The Thouless time(colored diamond) is determined as SFF at finite temperature is close enough to the analytical SFF(black dash line) of GOE at infinite temperature. It should be noticed that, by definition, the Thouless time is different from the dip time when the SFF reaches its minimum value. The Heisenberg time is shown as the black triangle marker. }
	\label{Scaling of g GOE}
\end{figure}

The SFF is another emerging diagnostic of quantum chaos~\cite{chaos_2017, RM2}, which measures the spectrum correlation of arbitrary two states. Specifically, we consider the following spectrum correlation,
\begin{align}
{\cal C}(E)=\frac{1}{L^{2}}\sum_{i,j}\delta\left[E-\left(E_{i}-E_{j}\right)\right],
\end{align}
It is evident that as long as the energy difference between two states is $E$, they contribute to ${\cal C}(E)$, regardless of whether they are the nearest neighbors or not. The SFF is obtained by a Fourier transformation as follows,
\begin{align}
\label{SFF_def}
K(t)=L\int dE e^{-iE t}{\cal C}(E)=\frac{1}{L}\left|{\rm Tr}\left({\rm e}^{-i\hat{H}t}\right)\right|^{2}.
\end{align}
It has been demonstrated that the ``dip-ramp-plateau" structure of SFF is a universal behavior for quantum chaotic systems. In the long time limit, $K(t)\to1$ according to our definition.  Moreover, the SFF has also been generalized to finite temperature by the analytical extension $t\to t-i\beta$, where $\beta=1/(k_{B}T)$~\cite{chaos_2017}. The finite temperature effect of SFF has been investigated in paradigmatic quantum chaotic systems, and the ``dip-ramp-plateau'' structure survives, which is reproduced in Fig.~\ref{Scaling of g GOE}.

The SFF is characterized by two time scales ~\cite{THt1,Tht2,ergo1,ergo2}; one is the Heisenberg time $t_{\rm H}$; the other one is the Thouless time $t_{\rm Th}$. The former one is defined as $t_{\rm H}=2\pi/\Delta$ with $\Delta$ being the mean energy level spacing which exponentially decreases as the system size increases~\cite{ergo1,ergo2}. In SFF, the Heisenberg time can be pinned down by the presence of plateau. The Thouless time is the timescale that characterizes how initial local information diffuses throughout the system, and is defined as the timescale beyond which the SFF becomes consistent the universal one of GOE.  In contrast to the Heisenberg time, the Thouless time increases polynomially as the system size increases~\cite{ergo1,ergo2,Tht3}.
In RMT, the analytical form of $K(t)$ at infinite temperature is available. For instance, $K(t)$ of GOE at intermediate and long time is written as~\cite{MSFF2020}
\begin{align}
K_{\rm a}(t)=\left\{
\begin{aligned}
&\frac{2t}{t_{\rm H}}-\frac{t}{t_{\rm H}}{\rm ln}\left(1+\frac{2t}{t_{\rm H}}\right) , 0<t<t_{\rm H}\\
&2-\frac{t}{t_{\rm H}}{\rm ln}\left(\frac{2t+t_{\rm H}}{2t-t_{\rm H}}\right),t>t_{\rm H}
\end{aligned}
\right.,
\label{Eq.AnalyGOE}
\end{align}
where we have taken the random average with respect to the probability of GOE. In the long time limit $K_{\rm a}(t)\to1$, which is consistent with our definition in Eq.(\ref{SFF_def}). 


With the Heisenberg and Thouless time, a quantity to measure the ergodicity in quantum systems is defined,
\begin{align}
g=\log_{10}(t_{\rm H}/t_{\rm Th})
\label{Eq.InErgo}
\end{align}
which is termed as ergodicity indicator~\cite{ergo1,ergo2}. In the thermodynamic limit, the ergodicity indicator $g$ describes a transition from the quantum ergodic regime $g\to \infty$ ($t_{\rm H}/t_{\rm Th}\to\infty$) to nonergodic regime $g\to -\infty$ ($t_{\rm H}/t_{\rm Th}\to0$). 

Similar to the generalization of SFF, we extend the ergodicity indicator to finite temperature since the Heisenberg and Thouless time can still be extracted from the SFF at finite temperature. In Fig.~\ref{Scaling of g GOE}, we illustrate the extraction of Heisenberg and Thouless time from SFF at finite temperature. Specifically, we numerically generate 150 samples of GOE at system size of $L=4096$ for each considered temperature. 
Following the method in \cite{ergo1,ergo2}, we apply the unfolding procedure, which maps each eigenvalue $E_i$ to a dimensionless number $\epsilon_i$ times average level spacing $\Delta$ and avoids the global energy dependence. Meanwhile, the information in local energy fluctuations is preserved~\cite{QCS2010}. Then the SFF after the unfolding procedure is written as
\begin{align}
K(t) =\frac{1}{L}\left|\sum_i{\rm e}^{-iE_i t}\right|^2\to& \frac{1}{L}\left|\sum_i {\rm e}^{-i(\epsilon_i\Delta) t}\right|^2\nonumber\\
=&\frac{1}{L}\left|\sum_i {\rm e}^{-i2\pi\epsilon_i t/t_{\rm H}}\right|^2.
\end{align}
In the last step, we use the definition of  the Heisenberg time $t_{\rm H} = 2\pi / \Delta$.
Therefore, the Heisenberg time, labelled by the black triangle in Fig.~\ref{Scaling of g GOE}, naturally serves as the time unit. The colored solid lines represent the average SFF of 150 samples in GOE at various temperature. It is evident that the SFF at finite temperature adheres to the analytical SFF $K_{\rm a}(t)$(black dash line) after the Thouless time(colored diamond). Numerically, the Thouless time is determined  when  finite temperature SFFs are sufficiently close to analytical SFF $K_{\rm a}(t)$, the quantitative criterion of which reads $\left|{\rm log}_{10}\left[\frac{K(t_{\rm Th})}{K_{\rm a}(t_{\rm Th})}\right]\right| < 0.05$ \cite{ergo1}. It is evident that a system with lower temperature exhibits a longer Thouless time, resulting in a smaller ergodicity indicator. This observation aligns with our intuition, indicating that the reduction in temperature suppresses the characteristics of chaos, as similarly observed in the previous section for Lanczos coefficients.

\begin{figure}[h]
	\centering
	\includegraphics[width=0.48\textwidth]{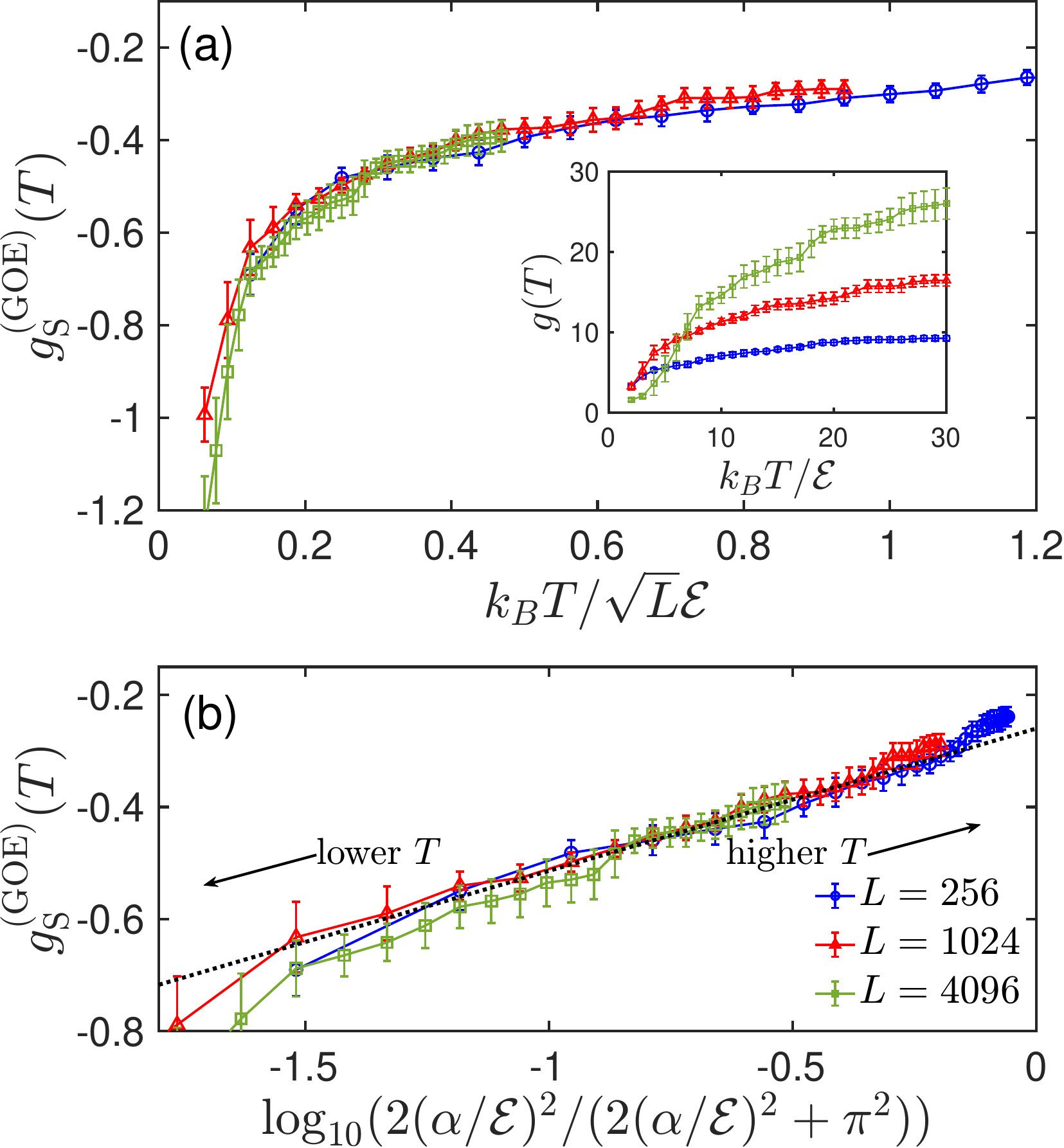}
	\caption{The scaled ergodicity indicator $g^{\rm (GOE)}_S(T)$ and its relation to the slope $\alpha(T)$ of Lanczos coefficients. (a): The initial ergodicity indicators displayed in the inset are different by a size dependence. After scaling, the scaled ergodicity indicators of different system size show collapse to a single curve.  (b): The linear relation between scaled ergodicity indicator and Lanczos coefficients slope. The error bar of scaled ergodicity indicator is the standard deviation of 10 sets of samples. The fluctuation of $\alpha(T)$ is so small during the sampling procedure that the error bar on Krylov complexity side is not shown. The black dotted line is plotted to show linearity.}
	\label{Fig3}
\end{figure}

\section{Scaling relation of GOE}
As of now, we present two chaos diagnostic tools that depend on temperature. A natural question arises: what is the relation between them?
In this section, we present a scaling relationship between the ergodicity indicator $g(T)$ 
and the slope of Lanczos coefficients $\alpha(T)$ at different temperature, which is showcased in Fig.~\ref{Fig3}(b). 
To this end, we firstly study the temperature dependence of ergodicity indicator $g(T)$, as illustrated in the inset of Fig.~\ref{Fig3}(a). The ergodicity indicator increases as the temperature ramps, as expected, and it also shows size dependency, the same as SFF~\cite{chaos_2017}. This size dependence precludes a meaningful and direct quantitative comparison between $g(T)$ and $\alpha(T)$.

In order to eliminate the size-dependent effects,  a scaled ergodicity indicator denoted as $g^{\rm (GOE)}_{\rm S}(T)$ has been proposed. It remains invariant with respect to different system size. Notably, for a particular system that can be simulated by GOE, the spectrum spans approximately from $-\mathcal{E}\sqrt{L}$ to $\mathcal{E}\sqrt{L}$. We postulate that temperature should be scaled by a factor of $\sqrt{L}$. Using the definition of Heisenberg time, we find that it is system size dependent, i.e., $t_{\rm H} = 2\pi L/ (E_{\rm max}-E_{\rm min})\sim2\pi \sqrt{L}/\mathcal{E}$. This contrasts with the Thouless time. As such, we can shift the value of the ergodicity indicator by amount of $\log_{10}(\sqrt{L})$, and define scaled ergodicity indicator as
\begin{equation}
\label{def_g}
	g^{\rm (GOE)}_{\rm S}(T) = {\rm log}_{10}\left[\frac{t_{\rm H}}{t_{\rm Th}(T)\sqrt{L}}\right].
\end{equation}
In Fig.~\ref{Fig3}(a), we present the scaled ergodicity indicator for different system size. 
Our results show that, within a broad temperature range, the scaled ergodicity indicators, denoted as $g^{\rm (GOE)}_{\rm S}(T)$, collapse to a single curve across various system sizes. 
We also would like to mention that the temperature in  Fig.~\ref{Fig3}(a) has also be scaled by $\sqrt{L}$, which confirms our previous postulation. The error bars represent the standard deviations of scaled ergodicity indicators from ten repetitions of sampling procedure.

Then the relation between $g^{\rm (GOE)}_{\rm S}(T)$ and  $\alpha(T)$ can be well defined. It has been demonstrated that the ratio of the time it takes for the SFF to reach the minimum and the plateau is proportional to $\sqrt{L}/(1+\beta ^2\mathcal{E}^2/2)$ for GUE~\cite{chaos_2017}. Noticing that the SFF at Thouless time is close to the minimum and the Heisenberg time is pinned down by the presence of plateau, we conjecture the ratio of Heisenberg time and Thouless time for GOE has similar form and can be written as
\begin{align}
\label{tt_ratio}
\frac{t_{\rm H}}{t_{\rm Th}} \propto\sqrt{L}\left(\frac{1}{1+\beta ^2\mathcal{E}^2/2}\right)^\delta.
\end{align}
Here, $\delta$ is a fitting parameter and $\beta\mathcal{E}$ is dimensionless.  
Upon substituting Eq.(\ref{tt_ratio}) and (\ref{Eq,alpha-T}) into Eq.(\ref{def_g}), we obtain a scaling relation between $g^{\rm (GOE)}_{\rm S}(T)$ and  $\alpha(T)$,
\begin{equation}
\label{gs-alpha}
	g^{\rm (GOE)}_{\rm S}(T) = \delta {\rm log}_{10}\left[\frac{2(\alpha(T)/\mathcal{E})^2}{2(\alpha(T)/\mathcal{E})^2+\pi^2}\right]+C,
\end{equation}
where we have used $\beta = 1/k_{\rm B}T$ and $C$ is a  fitting parameter. 
In Fig. \ref{Fig3}(b), we show that  Eq.(\ref{gs-alpha}) indeed holds around the temperature range of $k_{\rm B}T/{\cal E}\in[0.1,2]$. For different system sizes, the curves collapse to a single one.  By fitting Eq.(\ref{gs-alpha}) to numerical results, we find $\delta=0.25$ and $C = -0.26$. 
The origin of the error bars is the same as that for scaled ergodicity indicators.

\section{Scaling relation of random spin model}

Up to now, we work on the random matrices ensemble which describes the sufficiently complicated and chaotic systems. In this section, we study a 1D Ising  model in transverse and longitudinal field, TL-Ising model in short. In our calculation, the longitudinal field takes a random value on each site. This model is nonintegrable and chaotic at uniform transverse and random longitudinal field~\cite{Noh, RIsing1,Krylov11}. 
And the energy spectrum follows the Wigner-Dyson distribution, which is a hallmark of quantum chaos.
Besides, we choose the random parameter far away from the many-body-localization phase~\cite{Krylov11,RIsingL} to preserve the chaotic characteristics. In this system, the linear scaling relation is tested.

Specifically, the Hamiltonian of TL-Ising model is written as
\begin{equation}
	\hat H_{\rm spin} = \sum_i^N( -J\hat\sigma^x_i\hat\sigma^x_{i+1}+h\hat\sigma^z_i+g_i\hat\sigma^x_i),
	\label{EqRSM-Hamiltonian}
\end{equation}
where $J$ is the near-neighbor coupling and taken as the energy unite in the following discussion, $h$ represents the uniform transverse field and $g_i$ represents the site-dependent longitudinal field, $N$ is the total number of spin and $\hat\sigma_i^a$ with $a = x,z$ is the Pauli operator on $i$-th site. Formally, this operator should be considered in version of direct product that $\hat\sigma_i^a = ...\otimes\underbrace{\hat{\mathbb I}_2}_{i-1\text{-th site}}\otimes\underbrace{\hat\sigma^a}_{i\text{-th site}}\otimes\underbrace{\hat{\mathbb I}_2}_{i+1\text{-th site}}\otimes... $ with $\hat{\mathbb I}_2$ being the $2\times2$ identity operator. When the longitudinal field is turned off ($g_i = 0$ for each site), this model can be mapped to a free fermion model and  becomes integrable. As shown in \cite{Noh}, the integrability is broken in the presence of  uniform longitudinal field and the system tends to be chaotic. In our case, we take site-dependent longitudinal field $g_i$, which is sampled from a uniform distribution. 
\begin{figure}[t]
	\centering
	\includegraphics[width=0.48\textwidth]{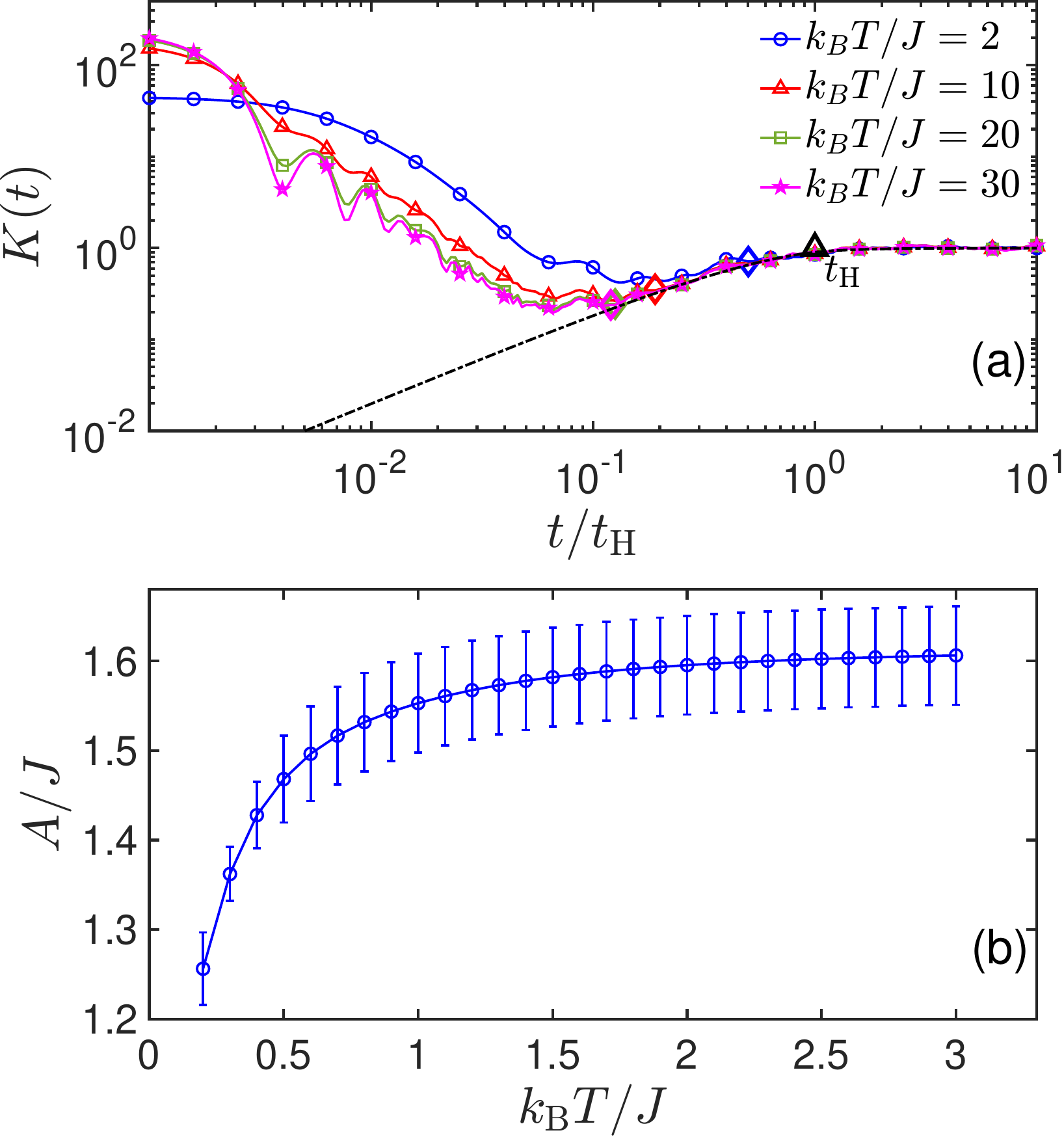}
	\caption{The SFF and logarithmic corrected slope $A(T)$ of spin model in transverse and random longitudinal field(TL-Ising model) at finite temperature. (a): The black dash line is the analytical SFF of GOE at infinite temperature. The colored solid lines represent the average SFF of TL-Ising model at finite temperature, from which the Thouless time is extracted(colored diamond). The Heisenberg time is still labelled by black triangle.  (b): The slope $A(T)$ of Lanczos coefficient defined by the Wightman inner product exhibits a progressive increase with rising temperature and saturates at high temperature. The error bar is the standard deviation of 10 sets of samples. }
	\label{Fig4}
\end{figure}

\begin{figure}[t]
	\centering
	\includegraphics[width=0.48\textwidth]{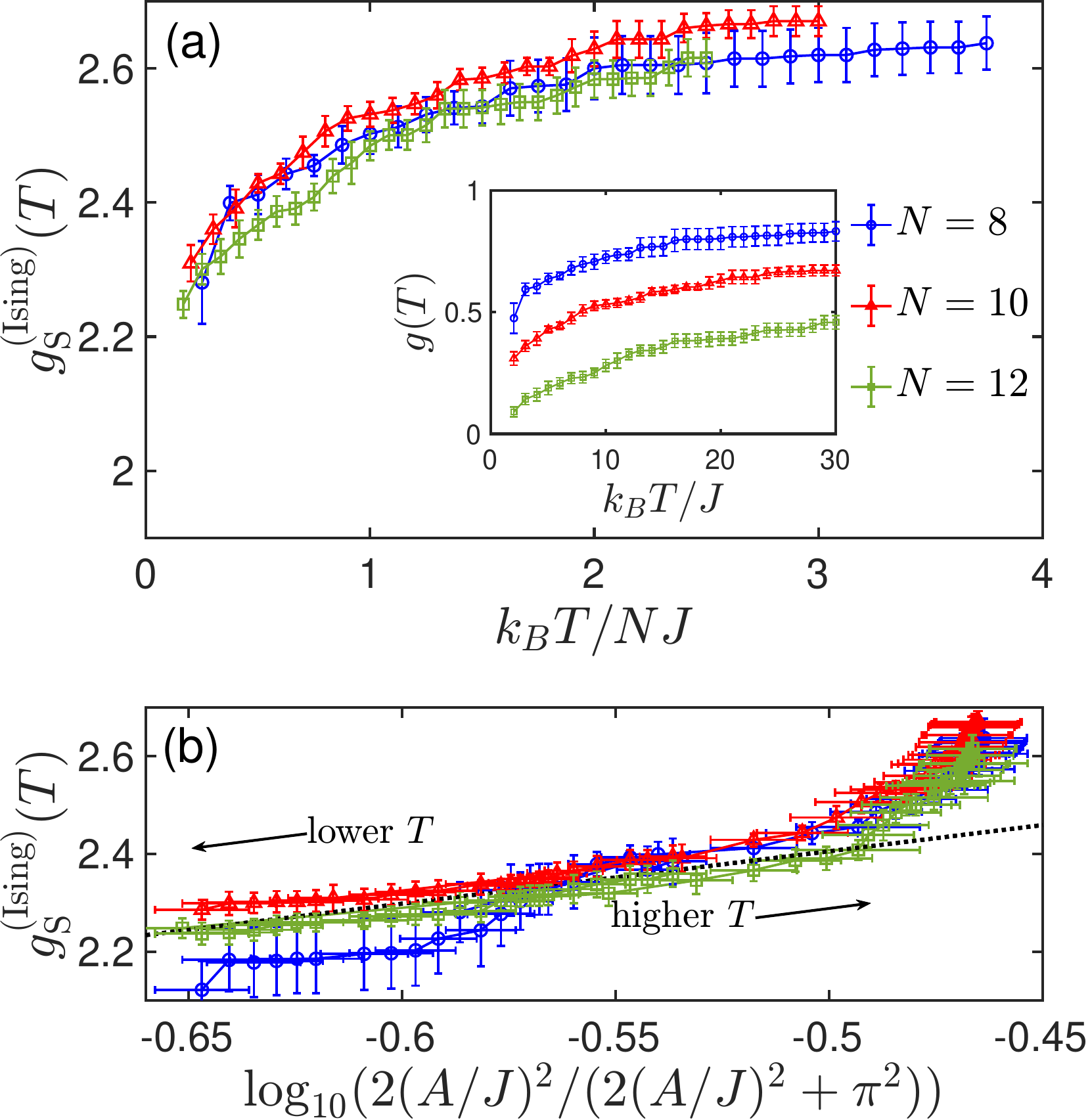}
	\caption{The scaled ergodicity indicator $g^{\rm (Ising)}_S(T)$ and the relation between $g^{\rm (Ising)}_S(T)$ and $A(T)$ for TL-Ising model. (a): The initial ergodicity indicator is displayed in the inset, which also presents system size dependence. The finite size dependence is eliminated in the scaled ergodicity indicator.  (b): The relation between the scaled ergodicity indicator and $A(T)$. The linearity can be observed at relatively  low temperature. The black dotted line is plotted to show linearity.}
	\label{Fig5}
\end{figure}

The SFF of TL-Ising model at finite temperature is shown in Fig.~\ref{Fig4}(a), where $N=8$, $h=J$ and $g_i$ is sampled from $[-J,J]$. Each colored curve of SFF data is taken average from 200 samples. The ``dip-ramp-plateau'' structure is observed, which confirms the existence of chaos. Using the same method illustrated in GOE, the Thouless time $t_{\rm Th}$ and Heisenberg time $t_{\rm H}$ are extracted from the SFF data. The Heisenberg time still serves as a time unit. At lower temperature, the SFF of TL-Ising model takes longer time to adhere to the analytical SFF of GOE, indicating larger Thouless time. 
Then the ergodicity indicator defined by Eq.(\ref{Eq.InErgo}) can be calculated. As expected, the ergodicity indicator shows a size dependence, seen in the inset of Fig.~\ref{Fig5}(a). As such, we need to scale the SFF to avoid the size effect.

The scaling method on TL-Ising model is different from the case of random matrix. For energy of TL-Ising model ranges from $\sim -JN$ to $\sim JN$. The temperature $k_BT/J$ is scaled to $k_BT/NJ$. In contrast to GOE, the Thouless time is proportion to $N^2$~\cite{ergo1,ergo2}. Therefore, the scaled ergodicity indicator for TL-Ising model is written as 
\begin{equation}
	g_{\rm S}^{\rm (Ising)}(T) = {\rm log}_{10}\left[\frac{t_{\rm H}N^2}{t_{\rm Th}(T)}\right].
\end{equation}
As shown in Fig.~\ref{Fig5}(a), the size effect on ergodicity indicator for TL-Ising model is almost eliminated. 

On the Krylov complexity side, we also use the Wightman inner product in the calculation of Lanczos coefficient $b_{n}$. 
In our calculation, the operator is chosen as $\hat{\sigma}^{(1)}_{z}\otimes\Pi_{i=2}^{N}\hat{\mathbb I}_2^{(i)}$. 
As the system is one dimensional, the linear growth of $b_n$ is corrected by logarithmic correction $b_n = A(T)n/{\rm ln}(n)$. Here $A(T)$ is the slope of $\left\lbrace b_n\right\rbrace $ sequence after logarithmic correction~\cite{Ehud}.  Considering  the disordered longitudinal field, we obtain the final $b_n$ by the mean value of 10 samples, and fit the results by a linear function. 
The temperature dependence of $A(T)$ is shown in Fig.~\ref{Fig4}(b). The points in figure represent the average of $A(T)$ from five repetitions and the error bars represent their standard deviations. It is obvious that $A(T)$ tends to saturate at high temperature limit, which indicates the TL-Ising model is chaotic but not sufficiently chaotic.

Then the scaling relation between the scaled ergodicity indicator and the Lanczos coefficients slope is established. Inspired by the similarity of spectral properties between GOE and TL-Ising model~\cite{ergo1,ergo2}, we expect that the scaling law established in  GOE should also hold for the TL-Ising model.
Therefore, the counterpart of Eq.~(\ref{gs-alpha}) can be written as
\begin{equation}
\label{fitspin}
   g_{\rm S}^{\rm (Ising)}(T) = \delta{\rm log}_{10}\left[\frac{2(A(T)/J)^2}{2(A(T)/J)^2+\pi^2}\right]+C.
\end{equation}
As shown in Fig.\ref{Fig5}(b), for different system size, the linear scaling relation appears at relatively low temperature, which is consistent with the result of GOE. By fitting the numerical results to Eq.(\ref{fitspin}), we find $\delta =1.07$ and $C=2.94$ for TL-Ising model. As temperature $k_{\rm B}T/J>0.8$, a deviation from linearity arises. We suspect that the deviation is attributed to the property that this model is not sufficiently chaotic, leading to a saturation of both the spectral form factor (SFF) and the slope of Lanczos coefficients at finite temperatures, as illustrated in Fig.\ref{Fig4}. One of the side effects on SFF of keeping increasing temperature is the emergence of oscillations~\cite{chaos_2017}, which causes the numerical SFF of Ising model to intersect with the analytical SFF of the GOE at an earlier time. Consequently, these oscillations further suppress the Thouless time and increase the value of scaled ergodicity indicator, which accounts for the observed deviation.

\section{conclusion}
In summary, we numerically investigate the finite-temperature behavior of K-complexity and SFF in the context of GOE and TL-Ising model. Within the GOE, the Lanczos coefficients $\left\lbrace b_n \right\rbrace$ defined by the Wightman inner product exhibit notable linear growth at finite temperatures, 
which is
not observed in the $\left\lbrace b_n \right\rbrace$ defined by the standard inner product. Utilizing the Wightman inner product, the slope $\alpha$ of $b_{n}$ saturates the universal upper bound $\pi k_BT$, aligning with the fact that the random matrix effectively simulates highly complex or chaotic systems. In the case of TL-Ising model, the behavior of $b_n$ defined by the Wightman inner product displays linear growth after a logarithmic correction. The slope $A$ converges to a specific value at infinite temperature, suggesting that the chaotic nature of this spin system is comparatively subdued compared to the GOE.

Furthermore, we extend the concept of the ergodicity indicator $g$ to finite temperatures. 
Through a scaling procedure, which relies on system characteristics, we eliminate the size-dependence and preserve information reflective of the chaotic features inherent in the system. 
Then, we establish a linear relationship between the scaled ergodicity indicator $g_{\rm S}(T)$ and 
the Lanczos coefficients slope $\alpha(T)$ or $A(T)$. 
Our results build a connection between these two aspects of quantum chaos diagnostic, despite their disparate origins.

\textit{Acknowledgement}.  This work is supported by Innovation Program for Quantum Science and Technology (Grants No.2021ZD0302001), NSFC (Grants No.12174300), the Fundamental Research Funds for the Central Universities (Grant No. 71211819000001) and Tang Scholar.

\textit{Note Added}. When finishing this manuscript, we notice a new preprint discussing Krylov complexity in RMT at finite temperature has also been proposed \cite{Temperature2}. The overlap part is consistent with ours.


\begin{thebibliography}{104}%
\makeatletter
\providecommand \@ifxundefined [1]{%
 \@ifx{#1\undefined}
}%
\providecommand \@ifnum [1]{%
 \ifnum #1\expandafter \@firstoftwo
 \else \expandafter \@secondoftwo
 \fi
}%
\providecommand \@ifx [1]{%
 \ifx #1\expandafter \@firstoftwo
 \else \expandafter \@secondoftwo
 \fi
}%
\providecommand \natexlab [1]{#1}%
\providecommand \enquote  [1]{#1}%
\providecommand \bibnamefont  [1]{#1}%
\providecommand \bibfnamefont [1]{#1}%
\providecommand \citenamefont [1]{#1}%
\providecommand \href@noop [0]{\@secondoftwo}%
\providecommand \href [0]{\begingroup \@sanitize@url \@href}%
\providecommand \@href[1]{\@@startlink{#1}\@@href}%
\providecommand \@@href[1]{\endgroup#1\@@endlink}%
\providecommand \@sanitize@url [0]{\catcode `\\12\catcode `\$12\catcode
  `\&12\catcode `\#12\catcode `\^12\catcode `\_12\catcode `\%12\relax}%
\providecommand \@@startlink[1]{}%
\providecommand \@@endlink[0]{}%
\providecommand \url  [0]{\begingroup\@sanitize@url \@url }%
\providecommand \@url [1]{\endgroup\@href {#1}{\urlprefix }}%
\providecommand \urlprefix  [0]{URL }%
\providecommand \Eprint [0]{\href }%
\providecommand \doibase [0]{https://dx.doi.org}%
\providecommand \selectlanguage [0]{\@gobble}%
\providecommand \bibinfo  [0]{\@secondoftwo}%
\providecommand \bibfield  [0]{\@secondoftwo}%
\providecommand \translation [1]{[#1]}%
\providecommand \BibitemOpen [0]{}%
\providecommand \bibitemStop [0]{}%
\providecommand \bibitemNoStop [0]{.\EOS\space}%
\providecommand \EOS [0]{\spacefactor3000\relax}%
\providecommand \BibitemShut  [1]{\csname bibitem#1\endcsname}%
\let\auto@bib@innerbib\@empty
\bibitem [{\citenamefont {Stanley}(1999)}]{SR5}%
  \BibitemOpen
  \bibfield  {author} {\bibinfo {author} {\bibfnamefont {H.~E.}\ \bibnamefont
  {Stanley}},\ }\bibfield  {title} {\bibinfo {title} {Scaling, universality,
  and renormalization: Three pillars of modern critical phenomena},\ }\href
  {\doibase/10.1103/RevModPhys.71.S358} {\bibfield  {journal} {\bibinfo
  {journal} {Rev. Mod. Phys.}\ }\textbf {\bibinfo {volume} {71}},\ \bibinfo
  {pages} {S358} (\bibinfo {year} {1999})}\BibitemShut {NoStop}%
\bibitem [{\citenamefont {Huckestein}(1995)}]{SR1}%
  \BibitemOpen
  \bibfield  {author} {\bibinfo {author} {\bibfnamefont {B.}~\bibnamefont
  {Huckestein}},\ }\bibfield  {title} {\bibinfo {title} {Scaling theory of the
  integer quantum hall effect},\ }\href {\doibase/10.1103/RevModPhys.67.357}
  {\bibfield  {journal} {\bibinfo  {journal} {Rev. Mod. Phys.}\ }\textbf
  {\bibinfo {volume} {67}},\ \bibinfo {pages} {357} (\bibinfo {year}
  {1995})}\BibitemShut {NoStop}%
\bibitem [{\citenamefont {Osterloh}\ \emph {et~al.}(2002)\citenamefont
  {Osterloh}, \citenamefont {Amico}, \citenamefont {Falci},\ and\ \citenamefont
  {Fazio}}]{SR2}%
  \BibitemOpen
  \bibfield  {author} {\bibinfo {author} {\bibfnamefont {A.}~\bibnamefont
  {Osterloh}}, \bibinfo {author} {\bibfnamefont {L.}~\bibnamefont {Amico}},
  \bibinfo {author} {\bibfnamefont {G.}~\bibnamefont {Falci}}, \ and\ \bibinfo
  {author} {\bibfnamefont {R.}~\bibnamefont {Fazio}},\ }\bibfield  {title}
  {\bibinfo {title} {Scaling of entanglement close to a quantum phase
  transition},\ }\href {\doibase/10.1038/416608a} {\bibfield  {journal}
  {\bibinfo  {journal} {Nature}\ }\textbf {\bibinfo {volume} {416}},\ \bibinfo
  {pages} {608} (\bibinfo {year} {2002})}\BibitemShut {NoStop}%
\bibitem [{\citenamefont {Garc\'{\i}a-Garc\'{\i}a}\ and\ \citenamefont
  {Wang}(2008)}]{SR3}%
  \BibitemOpen
  \bibfield  {author} {\bibinfo {author} {\bibfnamefont {A.~M.}\ \bibnamefont
  {Garc\'{\i}a-Garc\'{\i}a}}\ and\ \bibinfo {author} {\bibfnamefont
  {J.}~\bibnamefont {Wang}},\ }\bibfield  {title} {\bibinfo {title}
  {Universality in quantum chaos and the one-parameter scaling theory},\ }\href
  {\doibase/10.1103/PhysRevLett.100.070603} {\bibfield  {journal} {\bibinfo
  {journal} {Phys. Rev. Lett.}\ }\textbf {\bibinfo {volume} {100}},\ \bibinfo
  {pages} {070603} (\bibinfo {year} {2008})}\BibitemShut {NoStop}%
\bibitem [{\citenamefont {Serra}\ \emph {et~al.}(1998)\citenamefont {Serra},
  \citenamefont {Neirotti},\ and\ \citenamefont {Kais}}]{SR4}%
  \BibitemOpen
  \bibfield  {author} {\bibinfo {author} {\bibfnamefont {P.}~\bibnamefont
  {Serra}}, \bibinfo {author} {\bibfnamefont {J.~P.}\ \bibnamefont {Neirotti}},
  \ and\ \bibinfo {author} {\bibfnamefont {S.}~\bibnamefont {Kais}},\
  }\bibfield  {title} {\bibinfo {title} {Finite {Size} {Scaling} in {Quantum}
  {Mechanics}},\ }\href {\doibase/10.1021/jp9820572} {\bibfield  {journal}
  {\bibinfo  {journal} {The Journal of Physical Chemistry A}\ }\textbf
  {\bibinfo {volume} {102}},\ \bibinfo {pages} {9518} (\bibinfo {year}
  {1998})}\BibitemShut {NoStop}%
\bibitem [{\citenamefont {Pal}\ and\ \citenamefont {Huse}(2010)}]{MBLS1}%
  \BibitemOpen
  \bibfield  {author} {\bibinfo {author} {\bibfnamefont {A.}~\bibnamefont
  {Pal}}\ and\ \bibinfo {author} {\bibfnamefont {D.~A.}\ \bibnamefont {Huse}},\
  }\bibfield  {title} {\bibinfo {title} {Many-body localization phase
  transition},\ }\href {\doibase/10.1103/PhysRevB.82.174411} {\bibfield
  {journal} {\bibinfo  {journal} {Phys. Rev. B}\ }\textbf {\bibinfo {volume}
  {82}},\ \bibinfo {pages} {174411} (\bibinfo {year} {2010})}\BibitemShut
  {NoStop}%
\bibitem [{\citenamefont {Fisher}\ and\ \citenamefont {Berker}(1982)}]{PT1}%
  \BibitemOpen
  \bibfield  {author} {\bibinfo {author} {\bibfnamefont {M.~E.}\ \bibnamefont
  {Fisher}}\ and\ \bibinfo {author} {\bibfnamefont {A.~N.}\ \bibnamefont
  {Berker}},\ }\bibfield  {title} {\bibinfo {title} {Scaling for first-order
  phase transitions in thermodynamic and finite systems},\ }\href
  {\doibase/10.1103/PhysRevB.26.2507} {\bibfield  {journal} {\bibinfo
  {journal} {Phys. Rev. B}\ }\textbf {\bibinfo {volume} {26}},\ \bibinfo
  {pages} {2507} (\bibinfo {year} {1982})}\BibitemShut {NoStop}%
\bibitem [{\citenamefont {Hung}\ \emph {et~al.}(2011)\citenamefont {Hung},
  \citenamefont {Zhang}, \citenamefont {Gemelke},\ and\ \citenamefont
  {Chin}}]{SRnew1}%
  \BibitemOpen
  \bibfield  {author} {\bibinfo {author} {\bibfnamefont {C.-L.}\ \bibnamefont
  {Hung}}, \bibinfo {author} {\bibfnamefont {X.}~\bibnamefont {Zhang}},
  \bibinfo {author} {\bibfnamefont {N.}~\bibnamefont {Gemelke}}, \ and\
  \bibinfo {author} {\bibfnamefont {C.}~\bibnamefont {Chin}},\ }\bibfield
  {title} {\bibinfo {title} {Observation of scale invariance and universality
  in two-dimensional bose gases},\ }\href {\doibase/10.1038/nature09722}
  {\bibfield  {journal} {\bibinfo  {journal} {Nature}\ }\textbf {\bibinfo
  {volume} {470}},\ \bibinfo {pages} {236–239} (\bibinfo {year}
  {2011})}\BibitemShut {NoStop}%
\bibitem [{\citenamefont {Puebla}\ \emph {et~al.}(2019)\citenamefont {Puebla},
  \citenamefont {Marty},\ and\ \citenamefont {Plenio}}]{SRnew2}%
  \BibitemOpen
  \bibfield  {author} {\bibinfo {author} {\bibfnamefont {R.}~\bibnamefont
  {Puebla}}, \bibinfo {author} {\bibfnamefont {O.}~\bibnamefont {Marty}}, \
  and\ \bibinfo {author} {\bibfnamefont {M.~B.}\ \bibnamefont {Plenio}},\
  }\bibfield  {title} {\bibinfo {title} {Quantum kibble-zurek physics in
  long-range transverse-field ising models},\ }\href
  {\doibase/10.1103/PhysRevA.100.032115} {\bibfield  {journal} {\bibinfo
  {journal} {Phys. Rev. A}\ }\textbf {\bibinfo {volume} {100}},\ \bibinfo
  {pages} {032115} (\bibinfo {year} {2019})}\BibitemShut {NoStop}%
\bibitem [{\citenamefont {Chen}\ \emph {et~al.}(2019)\citenamefont {Chen},
  \citenamefont {Horikoshi}, \citenamefont {Yoshioka},\ and\ \citenamefont
  {Kuwata-Gonokami}}]{SRnew3}%
  \BibitemOpen
  \bibfield  {author} {\bibinfo {author} {\bibfnamefont {Y.}~\bibnamefont
  {Chen}}, \bibinfo {author} {\bibfnamefont {M.}~\bibnamefont {Horikoshi}},
  \bibinfo {author} {\bibfnamefont {K.}~\bibnamefont {Yoshioka}}, \ and\
  \bibinfo {author} {\bibfnamefont {M.}~\bibnamefont {Kuwata-Gonokami}},\
  }\bibfield  {title} {\bibinfo {title} {Dynamical critical behavior of an
  attractive bose-einstein condensate phase transition},\ }\href
  {\doibase/10.1103/PhysRevLett.122.040406} {\bibfield  {journal} {\bibinfo
  {journal} {Phys. Rev. Lett.}\ }\textbf {\bibinfo {volume} {122}},\ \bibinfo
  {pages} {040406} (\bibinfo {year} {2019})}\BibitemShut {NoStop}%
\bibitem [{\citenamefont {Chauhan}\ \emph {et~al.}(2022)\citenamefont
  {Chauhan}, \citenamefont {Kumar}, \citenamefont {Tiwari}, \citenamefont
  {Tiwari},\ and\ \citenamefont {Ghosh}}]{SRnew4}%
  \BibitemOpen
  \bibfield  {author} {\bibinfo {author} {\bibfnamefont {H.~C.}\ \bibnamefont
  {Chauhan}}, \bibinfo {author} {\bibfnamefont {B.}~\bibnamefont {Kumar}},
  \bibinfo {author} {\bibfnamefont {A.}~\bibnamefont {Tiwari}}, \bibinfo
  {author} {\bibfnamefont {J.~K.}\ \bibnamefont {Tiwari}}, \ and\ \bibinfo
  {author} {\bibfnamefont {S.}~\bibnamefont {Ghosh}},\ }\bibfield  {title}
  {\bibinfo {title} {Different critical exponents on two sides of a transition:
  Observation of crossover from ising to heisenberg exchange in skyrmion host
  ${\mathrm{cu}}_{2}{\mathrm{oseo}}_{3}$},\ }\href
  {\doibase/10.1103/PhysRevLett.128.015703} {\bibfield  {journal} {\bibinfo
  {journal} {Phys. Rev. Lett.}\ }\textbf {\bibinfo {volume} {128}},\ \bibinfo
  {pages} {015703} (\bibinfo {year} {2022})}\BibitemShut {NoStop}%
\bibitem [{\citenamefont {Sierant}\ \emph {et~al.}(2020)\citenamefont
  {Sierant}, \citenamefont {Delande},\ and\ \citenamefont
  {Zakrzewski}}]{ergo1}%
  \BibitemOpen
  \bibfield  {author} {\bibinfo {author} {\bibfnamefont {P.}~\bibnamefont
  {Sierant}}, \bibinfo {author} {\bibfnamefont {D.}~\bibnamefont {Delande}}, \
  and\ \bibinfo {author} {\bibfnamefont {J.}~\bibnamefont {Zakrzewski}},\
  }\bibfield  {title} {\bibinfo {title} {Thouless time analysis of anderson and
  many-body localization transitions},\ }\href
  {\doibase/10.1103/PhysRevLett.124.186601} {\bibfield  {journal} {\bibinfo
  {journal} {Phys. Rev. Lett.}\ }\textbf {\bibinfo {volume} {124}},\ \bibinfo
  {pages} {186601} (\bibinfo {year} {2020})}\BibitemShut {NoStop}%
\bibitem [{\citenamefont {\ifmmode~\check{S}\else \v{S}\fi{}untajs}\ \emph
  {et~al.}(2020)\citenamefont {\ifmmode~\check{S}\else \v{S}\fi{}untajs},
  \citenamefont {Bon\ifmmode~\check{c}\else \v{c}\fi{}a}, \citenamefont
  {Prosen},\ and\ \citenamefont {Vidmar}}]{ergo2}%
  \BibitemOpen
  \bibfield  {author} {\bibinfo {author} {\bibfnamefont {J.}~\bibnamefont
  {\ifmmode~\check{S}\else \v{S}\fi{}untajs}}, \bibinfo {author} {\bibfnamefont
  {J.}~\bibnamefont {Bon\ifmmode~\check{c}\else \v{c}\fi{}a}}, \bibinfo
  {author} {\bibfnamefont {T.~c.~v.}\ \bibnamefont {Prosen}}, \ and\ \bibinfo
  {author} {\bibfnamefont {L.}~\bibnamefont {Vidmar}},\ }\bibfield  {title}
  {\bibinfo {title} {Quantum chaos challenges many-body localization},\ }\href
  {\doibase/10.1103/PhysRevE.102.062144} {\bibfield  {journal} {\bibinfo
  {journal} {Phys. Rev. E}\ }\textbf {\bibinfo {volume} {102}},\ \bibinfo
  {pages} {062144} (\bibinfo {year} {2020})}\BibitemShut {NoStop}%
\bibitem [{\citenamefont {Cotler}\ \emph
  {et~al.}(2017{\natexlab{a}})\citenamefont {Cotler}, \citenamefont {Gur-Ari},
  \citenamefont {Hanada}, \citenamefont {Polchinski}, \citenamefont {Saad},
  \citenamefont {Shenker}, \citenamefont {Stanford}, \citenamefont
  {Streicher},\ and\ \citenamefont {Tezuka}}]{BH-SYK1}%
  \BibitemOpen
  \bibfield  {author} {\bibinfo {author} {\bibfnamefont {J.~S.}\ \bibnamefont
  {Cotler}}, \bibinfo {author} {\bibfnamefont {G.}~\bibnamefont {Gur-Ari}},
  \bibinfo {author} {\bibfnamefont {M.}~\bibnamefont {Hanada}}, \bibinfo
  {author} {\bibfnamefont {J.}~\bibnamefont {Polchinski}}, \bibinfo {author}
  {\bibfnamefont {P.}~\bibnamefont {Saad}}, \bibinfo {author} {\bibfnamefont
  {S.~H.}\ \bibnamefont {Shenker}}, \bibinfo {author} {\bibfnamefont
  {D.}~\bibnamefont {Stanford}}, \bibinfo {author} {\bibfnamefont
  {A.}~\bibnamefont {Streicher}}, \ and\ \bibinfo {author} {\bibfnamefont
  {M.}~\bibnamefont {Tezuka}},\ }\bibfield  {title} {\bibinfo {title} {Black
  holes and random matrices},\ }\href {\doibase/10.1007/JHEP05(2017)118}
  {\bibfield  {journal} {\bibinfo  {journal} {Journal of High Energy Physics}\
  }\textbf {\bibinfo {volume} {2017}},\ \bibinfo {pages} {118} (\bibinfo {year}
  {2017}{\natexlab{a}})}\BibitemShut {NoStop}%
\bibitem [{\citenamefont {Garc\'{\i}a-Garc\'{\i}a}\ and\ \citenamefont
  {Verbaarschot}(2016)}]{BH-SYK2}%
  \BibitemOpen
  \bibfield  {author} {\bibinfo {author} {\bibfnamefont {A.~M.}\ \bibnamefont
  {Garc\'{\i}a-Garc\'{\i}a}}\ and\ \bibinfo {author} {\bibfnamefont {J.~J.~M.}\
  \bibnamefont {Verbaarschot}},\ }\bibfield  {title} {\bibinfo {title}
  {Spectral and thermodynamic properties of the sachdev-ye-kitaev model},\
  }\href {\doibase/10.1103/PhysRevD.94.126010} {\bibfield  {journal} {\bibinfo
  {journal} {Phys. Rev. D}\ }\textbf {\bibinfo {volume} {94}},\ \bibinfo
  {pages} {126010} (\bibinfo {year} {2016})}\BibitemShut {NoStop}%
\bibitem [{\citenamefont {Cotler}\ \emph
  {et~al.}(2017{\natexlab{b}})\citenamefont {Cotler}, \citenamefont
  {Hunter-Jones}, \citenamefont {Liu},\ and\ \citenamefont
  {Yoshida}}]{chaos_2017}%
  \BibitemOpen
  \bibfield  {author} {\bibinfo {author} {\bibfnamefont {J.}~\bibnamefont
  {Cotler}}, \bibinfo {author} {\bibfnamefont {N.}~\bibnamefont
  {Hunter-Jones}}, \bibinfo {author} {\bibfnamefont {J.}~\bibnamefont {Liu}}, \
  and\ \bibinfo {author} {\bibfnamefont {B.}~\bibnamefont {Yoshida}},\
  }\bibfield  {title} {\bibinfo {title} {Chaos, complexity, and random
  matrices},\ }\href {\doibase/10.1007/JHEP11(2017)048} {\bibfield  {journal}
  {\bibinfo  {journal} {Journal of High Energy Physics}\ }\textbf {\bibinfo
  {volume} {2017}},\ \bibinfo {pages} {48} (\bibinfo {year}
  {2017}{\natexlab{b}})}\BibitemShut {NoStop}%
\bibitem [{\citenamefont {Deutsch}(2018)}]{ETH1}%
  \BibitemOpen
  \bibfield  {author} {\bibinfo {author} {\bibfnamefont {J.~M.}\ \bibnamefont
  {Deutsch}},\ }\bibfield  {title} {\bibinfo {title} {Eigenstate thermalization
  hypothesis},\ }\href {\doibase/10.1088/1361-6633/aac9f1} {\bibfield
  {journal} {\bibinfo  {journal} {Reports on Progress in Physics}\ }\textbf
  {\bibinfo {volume} {81}},\ \bibinfo {pages} {082001} (\bibinfo {year}
  {2018})}\BibitemShut {NoStop}%
\bibitem [{\citenamefont {Deutsch}(1991)}]{ETH2}%
  \BibitemOpen
  \bibfield  {author} {\bibinfo {author} {\bibfnamefont {J.~M.}\ \bibnamefont
  {Deutsch}},\ }\bibfield  {title} {\bibinfo {title} {Quantum statistical
  mechanics in a closed system},\ }\href {\doibase/10.1103/PhysRevA.43.2046}
  {\bibfield  {journal} {\bibinfo  {journal} {Phys. Rev. A}\ }\textbf {\bibinfo
  {volume} {43}},\ \bibinfo {pages} {2046} (\bibinfo {year}
  {1991})}\BibitemShut {NoStop}%
\bibitem [{\citenamefont {Srednicki}(1994)}]{ETH3}%
  \BibitemOpen
  \bibfield  {author} {\bibinfo {author} {\bibfnamefont {M.}~\bibnamefont
  {Srednicki}},\ }\bibfield  {title} {\bibinfo {title} {Chaos and quantum
  thermalization},\ }\href {\doibase/10.1103/PhysRevE.50.888} {\bibfield
  {journal} {\bibinfo  {journal} {Phys. Rev. E}\ }\textbf {\bibinfo {volume}
  {50}},\ \bibinfo {pages} {888} (\bibinfo {year} {1994})}\BibitemShut
  {NoStop}%
\bibitem [{\citenamefont {Rigol}\ \emph {et~al.}(2008)\citenamefont {Rigol},
  \citenamefont {Dunjko},\ and\ \citenamefont {Olshanii}}]{ETH4}%
  \BibitemOpen
  \bibfield  {author} {\bibinfo {author} {\bibfnamefont {M.}~\bibnamefont
  {Rigol}}, \bibinfo {author} {\bibfnamefont {V.}~\bibnamefont {Dunjko}}, \
  and\ \bibinfo {author} {\bibfnamefont {M.}~\bibnamefont {Olshanii}},\
  }\bibfield  {title} {\bibinfo {title} {Thermalization and its mechanism for
  generic isolated quantum systems},\ }\href {\doibase/10.1038/nature06838}
  {\bibfield  {journal} {\bibinfo  {journal} {Nature}\ }\textbf {\bibinfo
  {volume} {452}},\ \bibinfo {pages} {854} (\bibinfo {year}
  {2008})}\BibitemShut {NoStop}%
\bibitem [{\citenamefont {Kim}\ \emph {et~al.}(2014)\citenamefont {Kim},
  \citenamefont {Ikeda},\ and\ \citenamefont {Huse}}]{ETH5}%
  \BibitemOpen
  \bibfield  {author} {\bibinfo {author} {\bibfnamefont {H.}~\bibnamefont
  {Kim}}, \bibinfo {author} {\bibfnamefont {T.~N.}\ \bibnamefont {Ikeda}}, \
  and\ \bibinfo {author} {\bibfnamefont {D.~A.}\ \bibnamefont {Huse}},\
  }\bibfield  {title} {\bibinfo {title} {Testing whether all eigenstates obey
  the eigenstate thermalization hypothesis},\ }\href
  {\doibase/10.1103/PhysRevE.90.052105} {\bibfield  {journal} {\bibinfo
  {journal} {Phys. Rev. E}\ }\textbf {\bibinfo {volume} {90}},\ \bibinfo
  {pages} {052105} (\bibinfo {year} {2014})}\BibitemShut {NoStop}%
\bibitem [{\citenamefont {Abanin}\ \emph {et~al.}(2019)\citenamefont {Abanin},
  \citenamefont {Altman}, \citenamefont {Bloch},\ and\ \citenamefont
  {Serbyn}}]{ETH6}%
  \BibitemOpen
  \bibfield  {author} {\bibinfo {author} {\bibfnamefont {D.~A.}\ \bibnamefont
  {Abanin}}, \bibinfo {author} {\bibfnamefont {E.}~\bibnamefont {Altman}},
  \bibinfo {author} {\bibfnamefont {I.}~\bibnamefont {Bloch}}, \ and\ \bibinfo
  {author} {\bibfnamefont {M.}~\bibnamefont {Serbyn}},\ }\bibfield  {title}
  {\bibinfo {title} {Colloquium: Many-body localization, thermalization, and
  entanglement},\ }\href {\doibase/10.1103/RevModPhys.91.021001} {\bibfield
  {journal} {\bibinfo  {journal} {Rev. Mod. Phys.}\ }\textbf {\bibinfo {volume}
  {91}},\ \bibinfo {pages} {021001} (\bibinfo {year} {2019})}\BibitemShut
  {NoStop}%
\bibitem [{\citenamefont {Yan}\ \emph {et~al.}(2020)\citenamefont {Yan},
  \citenamefont {Cincio},\ and\ \citenamefont {Zurek}}]{IS1}%
  \BibitemOpen
  \bibfield  {author} {\bibinfo {author} {\bibfnamefont {B.}~\bibnamefont
  {Yan}}, \bibinfo {author} {\bibfnamefont {L.}~\bibnamefont {Cincio}}, \ and\
  \bibinfo {author} {\bibfnamefont {W.~H.}\ \bibnamefont {Zurek}},\ }\bibfield
  {title} {\bibinfo {title} {Information scrambling and loschmidt echo},\
  }\href {\doibase/10.1103/PhysRevLett.124.160603} {\bibfield  {journal}
  {\bibinfo  {journal} {Phys. Rev. Lett.}\ }\textbf {\bibinfo {volume} {124}},\
  \bibinfo {pages} {160603} (\bibinfo {year} {2020})}\BibitemShut {NoStop}%
\bibitem [{\citenamefont {Shen}\ \emph {et~al.}(2020)\citenamefont {Shen},
  \citenamefont {Zhang}, \citenamefont {You},\ and\ \citenamefont
  {Zhai}}]{IS2}%
  \BibitemOpen
  \bibfield  {author} {\bibinfo {author} {\bibfnamefont {H.}~\bibnamefont
  {Shen}}, \bibinfo {author} {\bibfnamefont {P.}~\bibnamefont {Zhang}},
  \bibinfo {author} {\bibfnamefont {Y.-Z.}\ \bibnamefont {You}}, \ and\
  \bibinfo {author} {\bibfnamefont {H.}~\bibnamefont {Zhai}},\ }\bibfield
  {title} {\bibinfo {title} {Information scrambling in quantum neural
  networks},\ }\href {\doibase/10.1103/PhysRevLett.124.200504} {\bibfield
  {journal} {\bibinfo  {journal} {Phys. Rev. Lett.}\ }\textbf {\bibinfo
  {volume} {124}},\ \bibinfo {pages} {200504} (\bibinfo {year}
  {2020})}\BibitemShut {NoStop}%
\bibitem [{\citenamefont {et~al.}(2021)}]{IS3}%
  \BibitemOpen
  \bibfield  {author} {\bibinfo {author} {\bibfnamefont {X.~M.}\ \bibnamefont
  {et~al.}},\ }\bibfield  {title} {\bibinfo {title} {Information scrambling in
  quantum circuits},\ }\href {\doibase/10.1126/science.abg5029} {\bibfield
  {journal} {\bibinfo  {journal} {Science}\ }\textbf {\bibinfo {volume}
  {374}},\ \bibinfo {pages} {1479} (\bibinfo {year} {2021})}\BibitemShut
  {NoStop}%
\bibitem [{\citenamefont {Landsman}\ \emph {et~al.}(2019)\citenamefont
  {Landsman}, \citenamefont {Figgatt}, \citenamefont {Schuster}, \citenamefont
  {Linke}, \citenamefont {Yoshida}, \citenamefont {Yao},\ and\ \citenamefont
  {Monroe}}]{IS4}%
  \BibitemOpen
  \bibfield  {author} {\bibinfo {author} {\bibfnamefont {K.~A.}\ \bibnamefont
  {Landsman}}, \bibinfo {author} {\bibfnamefont {C.}~\bibnamefont {Figgatt}},
  \bibinfo {author} {\bibfnamefont {T.}~\bibnamefont {Schuster}}, \bibinfo
  {author} {\bibfnamefont {N.~M.}\ \bibnamefont {Linke}}, \bibinfo {author}
  {\bibfnamefont {B.}~\bibnamefont {Yoshida}}, \bibinfo {author} {\bibfnamefont
  {N.~Y.}\ \bibnamefont {Yao}}, \ and\ \bibinfo {author} {\bibfnamefont
  {C.}~\bibnamefont {Monroe}},\ }\bibfield  {title} {\bibinfo {title} {Verified
  quantum information scrambling},\ }\href {\doibase/10.1038/s41586-019-0952-6}
  {\bibfield  {journal} {\bibinfo  {journal} {Nature}\ }\textbf {\bibinfo
  {volume} {567}},\ \bibinfo {pages} {61} (\bibinfo {year} {2019})}\BibitemShut
  {NoStop}%
\bibitem [{\citenamefont {\ifmmode \check{Z}\else
  \v{Z}\fi{}nidari\ifmmode~\check{c}\else \v{c}\fi{}}\ \emph
  {et~al.}(2008)\citenamefont {\ifmmode \check{Z}\else
  \v{Z}\fi{}nidari\ifmmode~\check{c}\else \v{c}\fi{}}, \citenamefont {Prosen},\
  and\ \citenamefont {Prelov\ifmmode~\check{s}\else
  \v{s}\fi{}ek}}]{QEntangle1}%
  \BibitemOpen
  \bibfield  {author} {\bibinfo {author} {\bibfnamefont {M.}~\bibnamefont
  {\ifmmode \check{Z}\else \v{Z}\fi{}nidari\ifmmode~\check{c}\else
  \v{c}\fi{}}}, \bibinfo {author} {\bibfnamefont {T.~c.~v.}\ \bibnamefont
  {Prosen}}, \ and\ \bibinfo {author} {\bibfnamefont {P.}~\bibnamefont
  {Prelov\ifmmode~\check{s}\else \v{s}\fi{}ek}},\ }\bibfield  {title} {\bibinfo
  {title} {Many-body localization in the heisenberg $xxz$ magnet in a random
  field},\ }\href {\doibase/10.1103/PhysRevB.77.064426} {\bibfield  {journal}
  {\bibinfo  {journal} {Phys. Rev. B}\ }\textbf {\bibinfo {volume} {77}},\
  \bibinfo {pages} {064426} (\bibinfo {year} {2008})}\BibitemShut {NoStop}%
\bibitem [{\citenamefont {Bardarson}\ \emph {et~al.}(2012)\citenamefont
  {Bardarson}, \citenamefont {Pollmann},\ and\ \citenamefont
  {Moore}}]{QEntangle2}%
  \BibitemOpen
  \bibfield  {author} {\bibinfo {author} {\bibfnamefont {J.~H.}\ \bibnamefont
  {Bardarson}}, \bibinfo {author} {\bibfnamefont {F.}~\bibnamefont {Pollmann}},
  \ and\ \bibinfo {author} {\bibfnamefont {J.~E.}\ \bibnamefont {Moore}},\
  }\bibfield  {title} {\bibinfo {title} {Unbounded growth of entanglement in
  models of many-body localization},\ }\href
  {\doibase/10.1103/PhysRevLett.109.017202} {\bibfield  {journal} {\bibinfo
  {journal} {Phys. Rev. Lett.}\ }\textbf {\bibinfo {volume} {109}},\ \bibinfo
  {pages} {017202} (\bibinfo {year} {2012})}\BibitemShut {NoStop}%
\bibitem [{\citenamefont {Shklovskii}\ \emph {et~al.}(1993)\citenamefont
  {Shklovskii}, \citenamefont {Shapiro}, \citenamefont {Sears}, \citenamefont
  {Lambrianides},\ and\ \citenamefont {Shore}}]{ES2}%
  \BibitemOpen
  \bibfield  {author} {\bibinfo {author} {\bibfnamefont {B.~I.}\ \bibnamefont
  {Shklovskii}}, \bibinfo {author} {\bibfnamefont {B.}~\bibnamefont {Shapiro}},
  \bibinfo {author} {\bibfnamefont {B.~R.}\ \bibnamefont {Sears}}, \bibinfo
  {author} {\bibfnamefont {P.}~\bibnamefont {Lambrianides}}, \ and\ \bibinfo
  {author} {\bibfnamefont {H.~B.}\ \bibnamefont {Shore}},\ }\bibfield  {title}
  {\bibinfo {title} {Statistics of spectra of disordered systems near the
  metal-insulator transition},\ }\href {\doibase/10.1103/PhysRevB.47.11487}
  {\bibfield  {journal} {\bibinfo  {journal} {Phys. Rev. B}\ }\textbf {\bibinfo
  {volume} {47}},\ \bibinfo {pages} {11487} (\bibinfo {year}
  {1993})}\BibitemShut {NoStop}%
\bibitem [{\citenamefont {Frisch}\ \emph {et~al.}(2014)\citenamefont {Frisch},
  \citenamefont {Mark}, \citenamefont {Aikawa}, \citenamefont {Ferlaino},
  \citenamefont {Bohn}, \citenamefont {Makrides}, \citenamefont {Petrov},\ and\
  \citenamefont {Kotochigova}}]{ES1}%
  \BibitemOpen
  \bibfield  {author} {\bibinfo {author} {\bibfnamefont {A.}~\bibnamefont
  {Frisch}}, \bibinfo {author} {\bibfnamefont {M.}~\bibnamefont {Mark}},
  \bibinfo {author} {\bibfnamefont {K.}~\bibnamefont {Aikawa}}, \bibinfo
  {author} {\bibfnamefont {F.}~\bibnamefont {Ferlaino}}, \bibinfo {author}
  {\bibfnamefont {J.~L.}\ \bibnamefont {Bohn}}, \bibinfo {author}
  {\bibfnamefont {C.}~\bibnamefont {Makrides}}, \bibinfo {author}
  {\bibfnamefont {A.}~\bibnamefont {Petrov}}, \ and\ \bibinfo {author}
  {\bibfnamefont {S.}~\bibnamefont {Kotochigova}},\ }\bibfield  {title}
  {\bibinfo {title} {Quantum chaos in ultracold collisions of gas-phase erbium
  atoms},\ }\href {\doibase/10.1038/nature13137} {\bibfield  {journal}
  {\bibinfo  {journal} {Nature}\ }\textbf {\bibinfo {volume} {507}},\ \bibinfo
  {pages} {475} (\bibinfo {year} {2014})}\BibitemShut {NoStop}%
\bibitem [{\citenamefont {Liu}(2018)}]{ES3}%
  \BibitemOpen
  \bibfield  {author} {\bibinfo {author} {\bibfnamefont {J.}~\bibnamefont
  {Liu}},\ }\bibfield  {title} {\bibinfo {title} {Spectral form factors and
  late time quantum chaos},\ }\href {\doibase/10.1103/PhysRevD.98.086026}
  {\bibfield  {journal} {\bibinfo  {journal} {Phys. Rev. D}\ }\textbf {\bibinfo
  {volume} {98}},\ \bibinfo {pages} {086026} (\bibinfo {year}
  {2018})}\BibitemShut {NoStop}%
\bibitem [{\citenamefont {von Keyserlingk}\ \emph {et~al.}(2018)\citenamefont
  {von Keyserlingk}, \citenamefont {Rakovszky}, \citenamefont {Pollmann},\ and\
  \citenamefont {Sondhi}}]{OTOC1}%
  \BibitemOpen
  \bibfield  {author} {\bibinfo {author} {\bibfnamefont {C.~W.}\ \bibnamefont
  {von Keyserlingk}}, \bibinfo {author} {\bibfnamefont {T.}~\bibnamefont
  {Rakovszky}}, \bibinfo {author} {\bibfnamefont {F.}~\bibnamefont {Pollmann}},
  \ and\ \bibinfo {author} {\bibfnamefont {S.~L.}\ \bibnamefont {Sondhi}},\
  }\bibfield  {title} {\bibinfo {title} {Operator hydrodynamics, otocs, and
  entanglement growth in systems without conservation laws},\ }\href
  {\doibase/10.1103/PhysRevX.8.021013} {\bibfield  {journal} {\bibinfo
  {journal} {Phys. Rev. X}\ }\textbf {\bibinfo {volume} {8}},\ \bibinfo {pages}
  {021013} (\bibinfo {year} {2018})}\BibitemShut {NoStop}%
\bibitem [{\citenamefont {Rakovszky}\ \emph {et~al.}(2018)\citenamefont
  {Rakovszky}, \citenamefont {Pollmann},\ and\ \citenamefont {von
  Keyserlingk}}]{OTOC3}%
  \BibitemOpen
  \bibfield  {author} {\bibinfo {author} {\bibfnamefont {T.}~\bibnamefont
  {Rakovszky}}, \bibinfo {author} {\bibfnamefont {F.}~\bibnamefont {Pollmann}},
  \ and\ \bibinfo {author} {\bibfnamefont {C.~W.}\ \bibnamefont {von
  Keyserlingk}},\ }\bibfield  {title} {\bibinfo {title} {Diffusive
  hydrodynamics of out-of-time-ordered correlators with charge conservation},\
  }\href {\doibase/10.1103/PhysRevX.8.031058} {\bibfield  {journal} {\bibinfo
  {journal} {Phys. Rev. X}\ }\textbf {\bibinfo {volume} {8}},\ \bibinfo {pages}
  {031058} (\bibinfo {year} {2018})}\BibitemShut {NoStop}%
\bibitem [{\citenamefont {Prosen}(2007)}]{OC1}%
  \BibitemOpen
  \bibfield  {author} {\bibinfo {author} {\bibfnamefont {T.}~\bibnamefont
  {Prosen}},\ }\bibfield  {title} {\bibinfo {title} {Chaos and complexity of
  quantum motion},\ }\href {\doibase/10.1088/1751-8113/40/28/S02} {\bibfield
  {journal} {\bibinfo  {journal} {Journal of Physics A: Mathematical and
  Theoretical}\ }\textbf {\bibinfo {volume} {40}},\ \bibinfo {pages} {7881}
  (\bibinfo {year} {2007})}\BibitemShut {NoStop}%
\bibitem [{\citenamefont {Gharibyan}\ \emph {et~al.}(2018)\citenamefont
  {Gharibyan}, \citenamefont {Hanada}, \citenamefont {Shenker},\ and\
  \citenamefont {Tezuka}}]{DRP1}%
  \BibitemOpen
  \bibfield  {author} {\bibinfo {author} {\bibfnamefont {H.}~\bibnamefont
  {Gharibyan}}, \bibinfo {author} {\bibfnamefont {M.}~\bibnamefont {Hanada}},
  \bibinfo {author} {\bibfnamefont {S.~H.}\ \bibnamefont {Shenker}}, \ and\
  \bibinfo {author} {\bibfnamefont {M.}~\bibnamefont {Tezuka}},\ }\bibfield
  {title} {\bibinfo {title} {Onset of random matrix behavior in scrambling
  systems},\ }\href {\doibase/10.1007/JHEP07(2018)124} {\bibfield  {journal}
  {\bibinfo  {journal} {Journal of High Energy Physics}\ }\textbf {\bibinfo
  {volume} {2018}},\ \bibinfo {pages} {124} (\bibinfo {year}
  {2018})}\BibitemShut {NoStop}%
\bibitem [{\citenamefont {Chen}\ and\ \citenamefont {Ludwig}(2018)}]{SFF1}%
  \BibitemOpen
  \bibfield  {author} {\bibinfo {author} {\bibfnamefont {X.}~\bibnamefont
  {Chen}}\ and\ \bibinfo {author} {\bibfnamefont {A.~W.~W.}\ \bibnamefont
  {Ludwig}},\ }\bibfield  {title} {\bibinfo {title} {Universal spectral
  correlations in the chaotic wave function and the development of quantum
  chaos},\ }\href {\doibase/10.1103/PhysRevB.98.064309} {\bibfield  {journal}
  {\bibinfo  {journal} {Phys. Rev. B}\ }\textbf {\bibinfo {volume} {98}},\
  \bibinfo {pages} {064309} (\bibinfo {year} {2018})}\BibitemShut {NoStop}%
\bibitem [{\citenamefont {Chan}\ \emph
  {et~al.}(2018{\natexlab{a}})\citenamefont {Chan}, \citenamefont {De~Luca},\
  and\ \citenamefont {Chalker}}]{SFF2}%
  \BibitemOpen
  \bibfield  {author} {\bibinfo {author} {\bibfnamefont {A.}~\bibnamefont
  {Chan}}, \bibinfo {author} {\bibfnamefont {A.}~\bibnamefont {De~Luca}}, \
  and\ \bibinfo {author} {\bibfnamefont {J.~T.}\ \bibnamefont {Chalker}},\
  }\bibfield  {title} {\bibinfo {title} {Spectral statistics in spatially
  extended chaotic quantum many-body systems},\ }\href
  {\doibase/10.1103/PhysRevLett.121.060601} {\bibfield  {journal} {\bibinfo
  {journal} {Phys. Rev. Lett.}\ }\textbf {\bibinfo {volume} {121}},\ \bibinfo
  {pages} {060601} (\bibinfo {year} {2018}{\natexlab{a}})}\BibitemShut
  {NoStop}%
\bibitem [{\citenamefont {Parker}\ \emph {et~al.}(2019)\citenamefont {Parker},
  \citenamefont {Cao}, \citenamefont {Avdoshkin}, \citenamefont {Scaffidi},\
  and\ \citenamefont {Altman}}]{Ehud}%
  \BibitemOpen
  \bibfield  {author} {\bibinfo {author} {\bibfnamefont {D.~E.}\ \bibnamefont
  {Parker}}, \bibinfo {author} {\bibfnamefont {X.}~\bibnamefont {Cao}},
  \bibinfo {author} {\bibfnamefont {A.}~\bibnamefont {Avdoshkin}}, \bibinfo
  {author} {\bibfnamefont {T.}~\bibnamefont {Scaffidi}}, \ and\ \bibinfo
  {author} {\bibfnamefont {E.}~\bibnamefont {Altman}},\ }\bibfield  {title}
  {\bibinfo {title} {A universal operator growth hypothesis},\ }\href
  {\doibase/10.1103/PhysRevX.9.041017} {\bibfield  {journal} {\bibinfo
  {journal} {Phys. Rev. X}\ }\textbf {\bibinfo {volume} {9}},\ \bibinfo {pages}
  {041017} (\bibinfo {year} {2019})}\BibitemShut {NoStop}%
\bibitem [{\citenamefont {Guhr}\ \emph {et~al.}(1998)\citenamefont {Guhr},
  \citenamefont {Müller–Groeling},\ and\ \citenamefont
  {Weidenmüller}}]{WD1}%
  \BibitemOpen
  \bibfield  {author} {\bibinfo {author} {\bibfnamefont {T.}~\bibnamefont
  {Guhr}}, \bibinfo {author} {\bibfnamefont {A.}~\bibnamefont
  {Müller–Groeling}}, \ and\ \bibinfo {author} {\bibfnamefont {H.~A.}\
  \bibnamefont {Weidenmüller}},\ }\bibfield  {title} {\bibinfo {title}
  {Random-matrix theories in quantum physics: common concepts},\ }\href
  {\doibase/https://doi.org/10.1016/S0370-1573(97)00088-4} {\bibfield
  {journal} {\bibinfo  {journal} {Physics Reports}\ }\textbf {\bibinfo {volume}
  {299}},\ \bibinfo {pages} {189} (\bibinfo {year} {1998})}\BibitemShut
  {NoStop}%
\bibitem [{\citenamefont {Mehta}(1991)}]{RM}%
  \BibitemOpen
  \bibfield  {author} {\bibinfo {author} {\bibfnamefont {M.~L.}\ \bibnamefont
  {Mehta}},\ }\href@noop {} {\emph {\bibinfo {title} {Random Matrices}}}\
  (\bibinfo  {publisher} {Oxford University Press},\ \bibinfo {address}
  {Boston},\ \bibinfo {year} {1991})\BibitemShut {NoStop}%
\bibitem [{\citenamefont {Haake}(2010)}]{QCS2010}%
  \BibitemOpen
  \bibfield  {author} {\bibinfo {author} {\bibfnamefont {F.}~\bibnamefont
  {Haake}},\ }\href {\doibase/10.1007/978-3-642-05428-0} {\emph {\bibinfo
  {title} {Quantum Signatures of Chaos}}}\ (\bibinfo  {publisher} {Springer
  Berlin Heidelberg},\ \bibinfo {year} {2010})\BibitemShut {NoStop}%
\bibitem [{\citenamefont {Brody}\ \emph {et~al.}(1981)\citenamefont {Brody},
  \citenamefont {Flores}, \citenamefont {French}, \citenamefont {Mello},
  \citenamefont {Pandey},\ and\ \citenamefont {Wong}}]{RM1}%
  \BibitemOpen
  \bibfield  {author} {\bibinfo {author} {\bibfnamefont {T.~A.}\ \bibnamefont
  {Brody}}, \bibinfo {author} {\bibfnamefont {J.}~\bibnamefont {Flores}},
  \bibinfo {author} {\bibfnamefont {J.~B.}\ \bibnamefont {French}}, \bibinfo
  {author} {\bibfnamefont {P.~A.}\ \bibnamefont {Mello}}, \bibinfo {author}
  {\bibfnamefont {A.}~\bibnamefont {Pandey}}, \ and\ \bibinfo {author}
  {\bibfnamefont {S.~S.~M.}\ \bibnamefont {Wong}},\ }\bibfield  {title}
  {\bibinfo {title} {Random-matrix physics: spectrum and strength
  fluctuations},\ }\href {\doibase/10.1103/RevModPhys.53.385} {\bibfield
  {journal} {\bibinfo  {journal} {Rev. Mod. Phys.}\ }\textbf {\bibinfo {volume}
  {53}},\ \bibinfo {pages} {385} (\bibinfo {year} {1981})}\BibitemShut
  {NoStop}%
\bibitem [{\citenamefont {Kos}\ \emph {et~al.}(2018)\citenamefont {Kos},
  \citenamefont {Ljubotina},\ and\ \citenamefont {Prosen}}]{RM2}%
  \BibitemOpen
  \bibfield  {author} {\bibinfo {author} {\bibfnamefont {P.}~\bibnamefont
  {Kos}}, \bibinfo {author} {\bibfnamefont {M.}~\bibnamefont {Ljubotina}}, \
  and\ \bibinfo {author} {\bibfnamefont {T.~c.~v.}\ \bibnamefont {Prosen}},\
  }\bibfield  {title} {\bibinfo {title} {Many-body quantum chaos: Analytic
  connection to random matrix theory},\ }\href
  {\doibase/10.1103/PhysRevX.8.021062} {\bibfield  {journal} {\bibinfo
  {journal} {Phys. Rev. X}\ }\textbf {\bibinfo {volume} {8}},\ \bibinfo {pages}
  {021062} (\bibinfo {year} {2018})}\BibitemShut {NoStop}%
\bibitem [{\citenamefont {Zhou}\ \emph {et~al.}(2024)\citenamefont {Zhou},
  \citenamefont {Zhou},\ and\ \citenamefont {Zhang}}]{Hubbard}%
  \BibitemOpen
  \bibfield  {author} {\bibinfo {author} {\bibfnamefont {Y.-N.}\ \bibnamefont
  {Zhou}}, \bibinfo {author} {\bibfnamefont {T.-G.}\ \bibnamefont {Zhou}}, \
  and\ \bibinfo {author} {\bibfnamefont {P.}~\bibnamefont {Zhang}},\ }\bibfield
   {title} {\bibinfo {title} {General properties of the spectral form factor in
  open quantum systems},\ }\href {\doibase/10.1007/s11467-024-1406-7}
  {\bibfield  {journal} {\bibinfo  {journal} {FRONTIERS OF PHYSICS}\ }\textbf
  {\bibinfo {volume} {19}},\ \bibinfo {pages} {31202} (\bibinfo {year}
  {2024})}\BibitemShut {NoStop}%
\bibitem [{\citenamefont {Wittmann~W.}\ \emph {et~al.}(2022)\citenamefont
  {Wittmann~W.}, \citenamefont {Castro}, \citenamefont {Foerster},\ and\
  \citenamefont {Santos}}]{Hubbard2}%
  \BibitemOpen
  \bibfield  {author} {\bibinfo {author} {\bibfnamefont {K.}~\bibnamefont
  {Wittmann~W.}}, \bibinfo {author} {\bibfnamefont {E.~R.}\ \bibnamefont
  {Castro}}, \bibinfo {author} {\bibfnamefont {A.}~\bibnamefont {Foerster}}, \
  and\ \bibinfo {author} {\bibfnamefont {L.~F.}\ \bibnamefont {Santos}},\
  }\bibfield  {title} {\bibinfo {title} {Interacting bosons in a triple well:
  Preface of many-body quantum chaos},\ }\href
  {\doibase/10.1103/PhysRevE.105.034204} {\bibfield  {journal} {\bibinfo
  {journal} {Phys. Rev. E}\ }\textbf {\bibinfo {volume} {105}},\ \bibinfo
  {pages} {034204} (\bibinfo {year} {2022})}\BibitemShut {NoStop}%
\bibitem [{\citenamefont {Dağ}\ \emph {et~al.}(2023)\citenamefont {Dağ},
  \citenamefont {Mistakidis}, \citenamefont {Chan},\ and\ \citenamefont
  {Sadeghpour}}]{Hubbard1}%
  \BibitemOpen
  \bibfield  {author} {\bibinfo {author} {\bibfnamefont {C.~B.}\ \bibnamefont
  {Dağ}}, \bibinfo {author} {\bibfnamefont {S.~I.}\ \bibnamefont
  {Mistakidis}}, \bibinfo {author} {\bibfnamefont {A.}~\bibnamefont {Chan}}, \
  and\ \bibinfo {author} {\bibfnamefont {H.~R.}\ \bibnamefont {Sadeghpour}},\
  }\bibfield  {title} {\bibinfo {title} {Many-body quantum chaos in
  stroboscopically-driven cold atoms},\ }\href
  {\doibase/10.1038/s42005-023-01258-1} {\bibfield  {journal} {\bibinfo
  {journal} {Communications Physics}\ }\textbf {\bibinfo {volume} {6}}
  (\bibinfo {year} {2023}),\ 10.1038/s42005-023-01258-1}\BibitemShut {NoStop}%
\bibitem [{\citenamefont {Saad}\ \emph {et~al.}(2019)\citenamefont {Saad},
  \citenamefont {Shenker},\ and\ \citenamefont {Stanford}}]{DRP2}%
  \BibitemOpen
  \bibfield  {author} {\bibinfo {author} {\bibfnamefont {P.}~\bibnamefont
  {Saad}}, \bibinfo {author} {\bibfnamefont {S.~H.}\ \bibnamefont {Shenker}}, \
  and\ \bibinfo {author} {\bibfnamefont {D.}~\bibnamefont {Stanford}},\
  }\href@noop {} {\bibinfo {title} {A semiclassical ramp in syk and in
  gravity},\ } (\bibinfo {year} {2019}),\ \Eprint
  {https://arxiv.org/abs/1806.06840} {arXiv:1806.06840 [hep-th]} \BibitemShut
  {NoStop}%
\bibitem [{\citenamefont {Winer}\ \emph {et~al.}(2020)\citenamefont {Winer},
  \citenamefont {Jian},\ and\ \citenamefont {Swingle}}]{DRP3}%
  \BibitemOpen
  \bibfield  {author} {\bibinfo {author} {\bibfnamefont {M.}~\bibnamefont
  {Winer}}, \bibinfo {author} {\bibfnamefont {S.-K.}\ \bibnamefont {Jian}}, \
  and\ \bibinfo {author} {\bibfnamefont {B.}~\bibnamefont {Swingle}},\
  }\bibfield  {title} {\bibinfo {title} {Exponential ramp in the quadratic
  sachdev-ye-kitaev model},\ }\href {\doibase/10.1103/PhysRevLett.125.250602}
  {\bibfield  {journal} {\bibinfo  {journal} {Phys. Rev. Lett.}\ }\textbf
  {\bibinfo {volume} {125}},\ \bibinfo {pages} {250602} (\bibinfo {year}
  {2020})}\BibitemShut {NoStop}%
\bibitem [{\citenamefont {Wei}\ \emph {et~al.}(2024)\citenamefont {Wei},
  \citenamefont {Tan},\ and\ \citenamefont {Zhang}}]{GSFF1}%
  \BibitemOpen
  \bibfield  {author} {\bibinfo {author} {\bibfnamefont {Z.}~\bibnamefont
  {Wei}}, \bibinfo {author} {\bibfnamefont {C.}~\bibnamefont {Tan}}, \ and\
  \bibinfo {author} {\bibfnamefont {R.}~\bibnamefont {Zhang}},\ }\bibfield
  {title} {\bibinfo {title} {Generalized spectral form factor in random matrix
  theory},\ }\href {\doibase/10.1103/PhysRevE.109.064208} {\bibfield  {journal}
  {\bibinfo  {journal} {Phys. Rev. E}\ }\textbf {\bibinfo {volume} {109}},\
  \bibinfo {pages} {064208} (\bibinfo {year} {2024})}\BibitemShut {NoStop}%
\bibitem [{\citenamefont {Cardella}(2021)}]{GSFF2}%
  \BibitemOpen
  \bibfield  {author} {\bibinfo {author} {\bibfnamefont {M.~A.}\ \bibnamefont
  {Cardella}},\ }\href@noop {} {\bibinfo {title} {A late times approximation
  for the syk spectral form factor},\ } (\bibinfo {year} {2021}),\ \Eprint
  {https://arxiv.org/abs/2102.01653} {arXiv:2102.01653 [hep-th]} \BibitemShut
  {NoStop}%
\bibitem [{\citenamefont {Gross}\ and\ \citenamefont
  {Rosenhaus}(2017)}]{GSFF3}%
  \BibitemOpen
  \bibfield  {author} {\bibinfo {author} {\bibfnamefont {D.~J.}\ \bibnamefont
  {Gross}}\ and\ \bibinfo {author} {\bibfnamefont {V.}~\bibnamefont
  {Rosenhaus}},\ }\bibfield  {title} {\bibinfo {title} {All point correlation
  functions in {SYK}},\ }\href {\doibase/10.1007/JHEP12(2017)148} {\bibfield
  {journal} {\bibinfo  {journal} {Journal of High Energy Physics}\ }\textbf
  {\bibinfo {volume} {2017}},\ \bibinfo {pages} {148} (\bibinfo {year}
  {2017})}\BibitemShut {NoStop}%
\bibitem [{\citenamefont {Nahum}\ \emph {et~al.}(2018)\citenamefont {Nahum},
  \citenamefont {Vijay},\ and\ \citenamefont {Haah}}]{OG2}%
  \BibitemOpen
  \bibfield  {author} {\bibinfo {author} {\bibfnamefont {A.}~\bibnamefont
  {Nahum}}, \bibinfo {author} {\bibfnamefont {S.}~\bibnamefont {Vijay}}, \ and\
  \bibinfo {author} {\bibfnamefont {J.}~\bibnamefont {Haah}},\ }\bibfield
  {title} {\bibinfo {title} {Operator spreading in random unitary circuits},\
  }\href {\doibase/10.1103/PhysRevX.8.021014} {\bibfield  {journal} {\bibinfo
  {journal} {Phys. Rev. X}\ }\textbf {\bibinfo {volume} {8}},\ \bibinfo {pages}
  {021014} (\bibinfo {year} {2018})}\BibitemShut {NoStop}%
\bibitem [{\citenamefont {Khemani}\ \emph {et~al.}(2018)\citenamefont
  {Khemani}, \citenamefont {Vishwanath},\ and\ \citenamefont {Huse}}]{OG4}%
  \BibitemOpen
  \bibfield  {author} {\bibinfo {author} {\bibfnamefont {V.}~\bibnamefont
  {Khemani}}, \bibinfo {author} {\bibfnamefont {A.}~\bibnamefont {Vishwanath}},
  \ and\ \bibinfo {author} {\bibfnamefont {D.~A.}\ \bibnamefont {Huse}},\
  }\bibfield  {title} {\bibinfo {title} {Operator spreading and the emergence
  of dissipative hydrodynamics under unitary evolution with conservation
  laws},\ }\href {\doibase/10.1103/PhysRevX.8.031057} {\bibfield  {journal}
  {\bibinfo  {journal} {Phys. Rev. X}\ }\textbf {\bibinfo {volume} {8}},\
  \bibinfo {pages} {031057} (\bibinfo {year} {2018})}\BibitemShut {NoStop}%
\bibitem [{\citenamefont {Chan}\ \emph
  {et~al.}(2018{\natexlab{b}})\citenamefont {Chan}, \citenamefont {De~Luca},\
  and\ \citenamefont {Chalker}}]{OG6}%
  \BibitemOpen
  \bibfield  {author} {\bibinfo {author} {\bibfnamefont {A.}~\bibnamefont
  {Chan}}, \bibinfo {author} {\bibfnamefont {A.}~\bibnamefont {De~Luca}}, \
  and\ \bibinfo {author} {\bibfnamefont {J.~T.}\ \bibnamefont {Chalker}},\
  }\bibfield  {title} {\bibinfo {title} {Solution of a minimal model for
  many-body quantum chaos},\ }\href {\doibase/10.1103/PhysRevX.8.041019}
  {\bibfield  {journal} {\bibinfo  {journal} {Phys. Rev. X}\ }\textbf {\bibinfo
  {volume} {8}},\ \bibinfo {pages} {041019} (\bibinfo {year}
  {2018}{\natexlab{b}})}\BibitemShut {NoStop}%
\bibitem [{\citenamefont {Roberts}\ \emph {et~al.}(2018)\citenamefont
  {Roberts}, \citenamefont {Stanford},\ and\ \citenamefont
  {Streicher}}]{operator-size}%
  \BibitemOpen
  \bibfield  {author} {\bibinfo {author} {\bibfnamefont {D.~A.}\ \bibnamefont
  {Roberts}}, \bibinfo {author} {\bibfnamefont {D.}~\bibnamefont {Stanford}}, \
  and\ \bibinfo {author} {\bibfnamefont {A.}~\bibnamefont {Streicher}},\
  }\bibfield  {title} {\bibinfo {title} {Operator growth in the syk model},\
  }\href {\doibase/10.1007/JHEP06(2018)122} {\bibfield  {journal} {\bibinfo
  {journal} {Journal of High Energy Physics}\ }\textbf {\bibinfo {volume}
  {2018}},\ \bibinfo {pages} {122} (\bibinfo {year} {2018})}\BibitemShut
  {NoStop}%
\bibitem [{\citenamefont {Barb{\'o}n}\ \emph {et~al.}(2020)\citenamefont
  {Barb{\'o}n}, \citenamefont {Mart{\'\i}n-Garc{\'\i}a},\ and\ \citenamefont
  {Sasieta}}]{Krylov6}%
  \BibitemOpen
  \bibfield  {author} {\bibinfo {author} {\bibfnamefont {J.~F.}\ \bibnamefont
  {Barb{\'o}n}}, \bibinfo {author} {\bibfnamefont {J.}~\bibnamefont
  {Mart{\'\i}n-Garc{\'\i}a}}, \ and\ \bibinfo {author} {\bibfnamefont
  {M.}~\bibnamefont {Sasieta}},\ }\bibfield  {title} {\bibinfo {title}
  {Momentum/complexity duality and the black hole interior},\ }\href
  {\doibase/10.1007/JHEP07(2020)169} {\bibfield  {journal} {\bibinfo  {journal}
  {Journal of High Energy Physics}\ }\textbf {\bibinfo {volume} {2020}},\
  \bibinfo {pages} {169} (\bibinfo {year} {2020})}\BibitemShut {NoStop}%
\bibitem [{\citenamefont {Mag{\'a}n}\ and\ \citenamefont
  {Sim{\'o}n}(2020)}]{Krylov7}%
  \BibitemOpen
  \bibfield  {author} {\bibinfo {author} {\bibfnamefont {J.~M.}\ \bibnamefont
  {Mag{\'a}n}}\ and\ \bibinfo {author} {\bibfnamefont {J.}~\bibnamefont
  {Sim{\'o}n}},\ }\bibfield  {title} {\bibinfo {title} {On operator growth and
  emergent poincar{\'e}symmetries},\ }\href {\doibase/10.1007/JHEP05(2020)071}
  {\bibfield  {journal} {\bibinfo  {journal} {Journal of High Energy Physics}\
  }\textbf {\bibinfo {volume} {2020}},\ \bibinfo {pages} {71} (\bibinfo {year}
  {2020})}\BibitemShut {NoStop}%
\bibitem [{\citenamefont {Kar}\ \emph {et~al.}(2022)\citenamefont {Kar},
  \citenamefont {Lamprou}, \citenamefont {Rozali},\ and\ \citenamefont
  {Sully}}]{Krylov12}%
  \BibitemOpen
  \bibfield  {author} {\bibinfo {author} {\bibfnamefont {A.}~\bibnamefont
  {Kar}}, \bibinfo {author} {\bibfnamefont {L.}~\bibnamefont {Lamprou}},
  \bibinfo {author} {\bibfnamefont {M.}~\bibnamefont {Rozali}}, \ and\ \bibinfo
  {author} {\bibfnamefont {J.}~\bibnamefont {Sully}},\ }\bibfield  {title}
  {\bibinfo {title} {Random matrix theory for complexity growth and black hole
  interiors},\ }\href {\doibase/10.1007/JHEP01(2022)016} {\bibfield  {journal}
  {\bibinfo  {journal} {Journal of High Energy Physics}\ }\textbf {\bibinfo
  {volume} {2022}},\ \bibinfo {pages} {16} (\bibinfo {year}
  {2022})}\BibitemShut {NoStop}%
\bibitem [{\citenamefont {Liu}\ \emph {et~al.}(2023)\citenamefont {Liu},
  \citenamefont {Tang},\ and\ \citenamefont {Zhai}}]{Hui_Krylov}%
  \BibitemOpen
  \bibfield  {author} {\bibinfo {author} {\bibfnamefont {C.}~\bibnamefont
  {Liu}}, \bibinfo {author} {\bibfnamefont {H.}~\bibnamefont {Tang}}, \ and\
  \bibinfo {author} {\bibfnamefont {H.}~\bibnamefont {Zhai}},\ }\bibfield
  {title} {\bibinfo {title} {Krylov complexity in open quantum systems},\
  }\href {\doibase/10.1103/PhysRevResearch.5.033085} {\bibfield  {journal}
  {\bibinfo  {journal} {Phys. Rev. Res.}\ }\textbf {\bibinfo {volume} {5}},\
  \bibinfo {pages} {033085} (\bibinfo {year} {2023})}\BibitemShut {NoStop}%
\bibitem [{\citenamefont {Bhattacharya}\ \emph {et~al.}(2022)\citenamefont
  {Bhattacharya}, \citenamefont {Nandy}, \citenamefont {Nath},\ and\
  \citenamefont {Sahu}}]{KC_OC1}%
  \BibitemOpen
  \bibfield  {author} {\bibinfo {author} {\bibfnamefont {A.}~\bibnamefont
  {Bhattacharya}}, \bibinfo {author} {\bibfnamefont {P.}~\bibnamefont {Nandy}},
  \bibinfo {author} {\bibfnamefont {P.~P.}\ \bibnamefont {Nath}}, \ and\
  \bibinfo {author} {\bibfnamefont {H.}~\bibnamefont {Sahu}},\ }\bibfield
  {title} {\bibinfo {title} {Operator growth and krylov construction in
  dissipative open quantum systems},\ }\href {\doibase/10.1007/jhep12(2022)081}
  {\bibfield  {journal} {\bibinfo  {journal} {Journal of High Energy Physics}\
  }\textbf {\bibinfo {volume} {2022}},\ \bibinfo {pages} {81} (\bibinfo {year}
  {2022})}\BibitemShut {NoStop}%
\bibitem [{\citenamefont {Bhattacharjee}\ \emph {et~al.}(2023)\citenamefont
  {Bhattacharjee}, \citenamefont {Cao}, \citenamefont {Nandy},\ and\
  \citenamefont {Pathak}}]{KC_OC2}%
  \BibitemOpen
  \bibfield  {author} {\bibinfo {author} {\bibfnamefont {B.}~\bibnamefont
  {Bhattacharjee}}, \bibinfo {author} {\bibfnamefont {X.}~\bibnamefont {Cao}},
  \bibinfo {author} {\bibfnamefont {P.}~\bibnamefont {Nandy}}, \ and\ \bibinfo
  {author} {\bibfnamefont {T.}~\bibnamefont {Pathak}},\ }\bibfield  {title}
  {\bibinfo {title} {Operator growth in open quantum systems: lessons from the
  dissipative syk},\ }\href {\doibase/10.1007/jhep03(2023)054} {\bibfield
  {journal} {\bibinfo  {journal} {Journal of High Energy Physics}\ }\textbf
  {\bibinfo {volume} {2023}},\ \bibinfo {pages} {054} (\bibinfo {year}
  {2023})}\BibitemShut {NoStop}%
\bibitem [{\citenamefont {Bhattacharya}\ \emph {et~al.}(2023)\citenamefont
  {Bhattacharya}, \citenamefont {Nandy}, \citenamefont {Nath},\ and\
  \citenamefont {Sahu}}]{KC_OC3}%
  \BibitemOpen
  \bibfield  {author} {\bibinfo {author} {\bibfnamefont {A.}~\bibnamefont
  {Bhattacharya}}, \bibinfo {author} {\bibfnamefont {P.}~\bibnamefont {Nandy}},
  \bibinfo {author} {\bibfnamefont {P.~P.}\ \bibnamefont {Nath}}, \ and\
  \bibinfo {author} {\bibfnamefont {H.}~\bibnamefont {Sahu}},\ }\bibfield
  {title} {\bibinfo {title} {On krylov complexity in open systems: an approach
  via bi-lanczos algorithm},\ }\href {\doibase/10.1007/jhep12(2023)066}
  {\bibfield  {journal} {\bibinfo  {journal} {Journal of High Energy Physics}\
  }\textbf {\bibinfo {volume} {2023}},\ \bibinfo {pages} {066} (\bibinfo {year}
  {2023})}\BibitemShut {NoStop}%
\bibitem [{\citenamefont {Bhattacharjee}\ \emph {et~al.}(2024)\citenamefont
  {Bhattacharjee}, \citenamefont {Nandy},\ and\ \citenamefont
  {Pathak}}]{KC_OC4}%
  \BibitemOpen
  \bibfield  {author} {\bibinfo {author} {\bibfnamefont {B.}~\bibnamefont
  {Bhattacharjee}}, \bibinfo {author} {\bibfnamefont {P.}~\bibnamefont
  {Nandy}}, \ and\ \bibinfo {author} {\bibfnamefont {T.}~\bibnamefont
  {Pathak}},\ }\bibfield  {title} {\bibinfo {title} {Operator dynamics in
  lindbladian syk: a krylov complexity perspective},\ }\href
  {\doibase/10.1007/jhep01(2024)094} {\bibfield  {journal} {\bibinfo  {journal}
  {Journal of High Energy Physics}\ }\textbf {\bibinfo {volume} {2024}},\
  \bibinfo {pages} {094} (\bibinfo {year} {2024})}\BibitemShut {NoStop}%
\bibitem [{\citenamefont {Barb{\'o}n}\ \emph {et~al.}(2019)\citenamefont
  {Barb{\'o}n}, \citenamefont {Rabinovici}, \citenamefont {Shir},\ and\
  \citenamefont {Sinha}}]{Krylov1}%
  \BibitemOpen
  \bibfield  {author} {\bibinfo {author} {\bibfnamefont {J.~L.~F.}\
  \bibnamefont {Barb{\'o}n}}, \bibinfo {author} {\bibfnamefont
  {E.}~\bibnamefont {Rabinovici}}, \bibinfo {author} {\bibfnamefont
  {R.}~\bibnamefont {Shir}}, \ and\ \bibinfo {author} {\bibfnamefont
  {R.}~\bibnamefont {Sinha}},\ }\bibfield  {title} {\bibinfo {title} {On the
  evolution of operator complexity beyond scrambling},\ }\href
  {\doibase/10.1007/JHEP10(2019)264} {\bibfield  {journal} {\bibinfo  {journal}
  {Journal of High Energy Physics}\ }\textbf {\bibinfo {volume} {2019}},\
  \bibinfo {pages} {264} (\bibinfo {year} {2019})}\BibitemShut {NoStop}%
\bibitem [{\citenamefont {Dymarsky}\ and\ \citenamefont
  {Gorsky}(2020)}]{Krylov2}%
  \BibitemOpen
  \bibfield  {author} {\bibinfo {author} {\bibfnamefont {A.}~\bibnamefont
  {Dymarsky}}\ and\ \bibinfo {author} {\bibfnamefont {A.}~\bibnamefont
  {Gorsky}},\ }\bibfield  {title} {\bibinfo {title} {Quantum chaos as
  delocalization in krylov space},\ }\href
  {\doibase/10.1103/PhysRevB.102.085137} {\bibfield  {journal} {\bibinfo
  {journal} {Phys. Rev. B}\ }\textbf {\bibinfo {volume} {102}},\ \bibinfo
  {pages} {085137} (\bibinfo {year} {2020})}\BibitemShut {NoStop}%
\bibitem [{\citenamefont {Xu}\ \emph {et~al.}(2020)\citenamefont {Xu},
  \citenamefont {Scaffidi},\ and\ \citenamefont {Cao}}]{Krylov3}%
  \BibitemOpen
  \bibfield  {author} {\bibinfo {author} {\bibfnamefont {T.}~\bibnamefont
  {Xu}}, \bibinfo {author} {\bibfnamefont {T.}~\bibnamefont {Scaffidi}}, \ and\
  \bibinfo {author} {\bibfnamefont {X.}~\bibnamefont {Cao}},\ }\bibfield
  {title} {\bibinfo {title} {Does scrambling equal chaos?}\ }\href
  {\doibase/10.1103/PhysRevLett.124.140602} {\bibfield  {journal} {\bibinfo
  {journal} {Phys. Rev. Lett.}\ }\textbf {\bibinfo {volume} {124}},\ \bibinfo
  {pages} {140602} (\bibinfo {year} {2020})}\BibitemShut {NoStop}%
\bibitem [{\citenamefont {Avdoshkin}\ and\ \citenamefont
  {Dymarsky}(2020)}]{Krylov4}%
  \BibitemOpen
  \bibfield  {author} {\bibinfo {author} {\bibfnamefont {A.}~\bibnamefont
  {Avdoshkin}}\ and\ \bibinfo {author} {\bibfnamefont {A.}~\bibnamefont
  {Dymarsky}},\ }\bibfield  {title} {\bibinfo {title} {Euclidean operator
  growth and quantum chaos},\ }\href
  {\doibase/10.1103/PhysRevResearch.2.043234} {\bibfield  {journal} {\bibinfo
  {journal} {Phys. Rev. Res.}\ }\textbf {\bibinfo {volume} {2}},\ \bibinfo
  {pages} {043234} (\bibinfo {year} {2020})}\BibitemShut {NoStop}%
\bibitem [{\citenamefont {Rabinovici}\ \emph {et~al.}(2021)\citenamefont
  {Rabinovici}, \citenamefont {S{\'a}nchez-Garrido}, \citenamefont {Shir},\
  and\ \citenamefont {Sonner}}]{Krylov9}%
  \BibitemOpen
  \bibfield  {author} {\bibinfo {author} {\bibfnamefont {E.}~\bibnamefont
  {Rabinovici}}, \bibinfo {author} {\bibfnamefont {A.}~\bibnamefont
  {S{\'a}nchez-Garrido}}, \bibinfo {author} {\bibfnamefont {R.}~\bibnamefont
  {Shir}}, \ and\ \bibinfo {author} {\bibfnamefont {J.}~\bibnamefont
  {Sonner}},\ }\bibfield  {title} {\bibinfo {title} {Operator complexity: a
  journey to the edge of krylov space},\ }\href
  {\doibase/10.1007/JHEP06(2021)062} {\bibfield  {journal} {\bibinfo  {journal}
  {Journal of High Energy Physics}\ }\textbf {\bibinfo {volume} {2021}},\
  \bibinfo {pages} {62} (\bibinfo {year} {2021})}\BibitemShut {NoStop}%
\bibitem [{\citenamefont {Trigueros}\ and\ \citenamefont
  {Lin}(2022)}]{Krylov11}%
  \BibitemOpen
  \bibfield  {author} {\bibinfo {author} {\bibfnamefont {F.~B.}\ \bibnamefont
  {Trigueros}}\ and\ \bibinfo {author} {\bibfnamefont {C.-J.}\ \bibnamefont
  {Lin}},\ }\bibfield  {title} {\bibinfo {title} {{Krylov complexity of
  many-body localization: Operator localization in Krylov basis}},\ }\href
  {\doibase/10.21468/SciPostPhys.13.2.037} {\bibfield  {journal} {\bibinfo
  {journal} {SciPost Phys.}\ }\textbf {\bibinfo {volume} {13}},\ \bibinfo
  {pages} {037} (\bibinfo {year} {2022})}\BibitemShut {NoStop}%
\bibitem [{\citenamefont {Kim}\ \emph {et~al.}(2022)\citenamefont {Kim},
  \citenamefont {Murugan}, \citenamefont {Olle},\ and\ \citenamefont
  {Rosa}}]{Krylov13}%
  \BibitemOpen
  \bibfield  {author} {\bibinfo {author} {\bibfnamefont {J.}~\bibnamefont
  {Kim}}, \bibinfo {author} {\bibfnamefont {J.}~\bibnamefont {Murugan}},
  \bibinfo {author} {\bibfnamefont {J.}~\bibnamefont {Olle}}, \ and\ \bibinfo
  {author} {\bibfnamefont {D.}~\bibnamefont {Rosa}},\ }\bibfield  {title}
  {\bibinfo {title} {Operator delocalization in quantum networks},\ }\href
  {\doibase/10.1103/PhysRevA.105.L010201} {\bibfield  {journal} {\bibinfo
  {journal} {Phys. Rev. A}\ }\textbf {\bibinfo {volume} {105}},\ \bibinfo
  {pages} {L010201} (\bibinfo {year} {2022})}\BibitemShut {NoStop}%
\bibitem [{\citenamefont {H{\"o}rnedal}\ \emph {et~al.}(2022)\citenamefont
  {H{\"o}rnedal}, \citenamefont {Carabba}, \citenamefont {Matsoukas-Roubeas},\
  and\ \citenamefont {del Campo}}]{Krylov14}%
  \BibitemOpen
  \bibfield  {author} {\bibinfo {author} {\bibfnamefont {N.}~\bibnamefont
  {H{\"o}rnedal}}, \bibinfo {author} {\bibfnamefont {N.}~\bibnamefont
  {Carabba}}, \bibinfo {author} {\bibfnamefont {A.~S.}\ \bibnamefont
  {Matsoukas-Roubeas}}, \ and\ \bibinfo {author} {\bibfnamefont
  {A.}~\bibnamefont {del Campo}},\ }\bibfield  {title} {\bibinfo {title}
  {Ultimate speed limits to the growth of operator complexity},\ }\href
  {\doibase/10.1038/s42005-022-00985-1} {\bibfield  {journal} {\bibinfo
  {journal} {Communications Physics}\ }\textbf {\bibinfo {volume} {5}},\
  \bibinfo {pages} {207} (\bibinfo {year} {2022})}\BibitemShut {NoStop}%
\bibitem [{\citenamefont {Rabinovici}\ \emph
  {et~al.}(2022{\natexlab{a}})\citenamefont {Rabinovici}, \citenamefont
  {S{\'a}nchez-Garrido}, \citenamefont {Shir},\ and\ \citenamefont
  {Sonner}}]{Krylov15}%
  \BibitemOpen
  \bibfield  {author} {\bibinfo {author} {\bibfnamefont {E.}~\bibnamefont
  {Rabinovici}}, \bibinfo {author} {\bibfnamefont {A.}~\bibnamefont
  {S{\'a}nchez-Garrido}}, \bibinfo {author} {\bibfnamefont {R.}~\bibnamefont
  {Shir}}, \ and\ \bibinfo {author} {\bibfnamefont {J.}~\bibnamefont
  {Sonner}},\ }\bibfield  {title} {\bibinfo {title} {Krylov localization and
  suppression of complexity},\ }\href {\doibase/10.1007/JHEP03(2022)211}
  {\bibfield  {journal} {\bibinfo  {journal} {Journal of High Energy Physics}\
  }\textbf {\bibinfo {volume} {2022}},\ \bibinfo {pages} {211} (\bibinfo {year}
  {2022}{\natexlab{a}})}\BibitemShut {NoStop}%
\bibitem [{\citenamefont {Bhattacharjee}\ \emph {et~al.}(2022)\citenamefont
  {Bhattacharjee}, \citenamefont {Cao}, \citenamefont {Nandy},\ and\
  \citenamefont {Pathak}}]{Krylov16}%
  \BibitemOpen
  \bibfield  {author} {\bibinfo {author} {\bibfnamefont {B.}~\bibnamefont
  {Bhattacharjee}}, \bibinfo {author} {\bibfnamefont {X.}~\bibnamefont {Cao}},
  \bibinfo {author} {\bibfnamefont {P.}~\bibnamefont {Nandy}}, \ and\ \bibinfo
  {author} {\bibfnamefont {T.}~\bibnamefont {Pathak}},\ }\bibfield  {title}
  {\bibinfo {title} {Krylov complexity in saddle-dominated scrambling},\ }\href
  {\doibase/10.1007/JHEP05(2022)174} {\bibfield  {journal} {\bibinfo  {journal}
  {Journal of High Energy Physics}\ }\textbf {\bibinfo {volume} {2022}},\
  \bibinfo {pages} {174} (\bibinfo {year} {2022})}\BibitemShut {NoStop}%
\bibitem [{\citenamefont {Heveling}\ \emph {et~al.}(2022)\citenamefont
  {Heveling}, \citenamefont {Wang},\ and\ \citenamefont {Gemmer}}]{Krylov18}%
  \BibitemOpen
  \bibfield  {author} {\bibinfo {author} {\bibfnamefont {R.}~\bibnamefont
  {Heveling}}, \bibinfo {author} {\bibfnamefont {J.}~\bibnamefont {Wang}}, \
  and\ \bibinfo {author} {\bibfnamefont {J.}~\bibnamefont {Gemmer}},\
  }\bibfield  {title} {\bibinfo {title} {Numerically probing the universal
  operator growth hypothesis},\ }\href {\doibase/10.1103/PhysRevE.106.014152}
  {\bibfield  {journal} {\bibinfo  {journal} {Phys. Rev. E}\ }\textbf {\bibinfo
  {volume} {106}},\ \bibinfo {pages} {014152} (\bibinfo {year}
  {2022})}\BibitemShut {NoStop}%
\bibitem [{\citenamefont {Adhikari}\ and\ \citenamefont
  {Choudhury}(2022)}]{Krylov19}%
  \BibitemOpen
  \bibfield  {author} {\bibinfo {author} {\bibfnamefont {K.}~\bibnamefont
  {Adhikari}}\ and\ \bibinfo {author} {\bibfnamefont {S.}~\bibnamefont
  {Choudhury}},\ }\bibfield  {title} {\bibinfo {title} {Cosmological krylov
  complexity},\ }\href {\doibase/10.1002/prop.202200126} {\bibfield  {journal}
  {\bibinfo  {journal} {Fortschritte der Physik}\ }\textbf {\bibinfo {volume}
  {70}},\ \bibinfo {pages} {2200126} (\bibinfo {year} {2022})}\BibitemShut
  {NoStop}%
\bibitem [{\citenamefont {Adhikari}\ \emph {et~al.}(2023)\citenamefont
  {Adhikari}, \citenamefont {Choudhury},\ and\ \citenamefont {Roy}}]{Krylov20}%
  \BibitemOpen
  \bibfield  {author} {\bibinfo {author} {\bibfnamefont {K.}~\bibnamefont
  {Adhikari}}, \bibinfo {author} {\bibfnamefont {S.}~\bibnamefont {Choudhury}},
  \ and\ \bibinfo {author} {\bibfnamefont {A.}~\bibnamefont {Roy}},\ }\bibfield
   {title} {\bibinfo {title} {Krylov complexity in quantum field theory},\
  }\href {\doibase/https://doi.org/10.1016/j.nuclphysb.2023.116263} {\bibfield
  {journal} {\bibinfo  {journal} {Nuclear Physics B}\ }\textbf {\bibinfo
  {volume} {993}},\ \bibinfo {pages} {116263} (\bibinfo {year}
  {2023})}\BibitemShut {NoStop}%
\bibitem [{\citenamefont {Caputa}\ and\ \citenamefont {Liu}(2022)}]{Krylov21}%
  \BibitemOpen
  \bibfield  {author} {\bibinfo {author} {\bibfnamefont {P.}~\bibnamefont
  {Caputa}}\ and\ \bibinfo {author} {\bibfnamefont {S.}~\bibnamefont {Liu}},\
  }\bibfield  {title} {\bibinfo {title} {Quantum complexity and topological
  phases of matter},\ }\href {\doibase/10.1103/PhysRevB.106.195125} {\bibfield
  {journal} {\bibinfo  {journal} {Phys. Rev. B}\ }\textbf {\bibinfo {volume}
  {106}},\ \bibinfo {pages} {195125} (\bibinfo {year} {2022})}\BibitemShut
  {NoStop}%
\bibitem [{\citenamefont {M{\"u}ck}\ and\ \citenamefont
  {Yang}(2022)}]{Krylov22}%
  \BibitemOpen
  \bibfield  {author} {\bibinfo {author} {\bibfnamefont {W.}~\bibnamefont
  {M{\"u}ck}}\ and\ \bibinfo {author} {\bibfnamefont {Y.}~\bibnamefont
  {Yang}},\ }\bibfield  {title} {\bibinfo {title} {Krylov complexity and
  orthogonal polynomials},\ }\href
  {\doibase/https://doi.org/10.1016/j.nuclphysb.2022.115948} {\bibfield
  {journal} {\bibinfo  {journal} {Nuclear Physics B}\ }\textbf {\bibinfo
  {volume} {984}},\ \bibinfo {pages} {115948} (\bibinfo {year}
  {2022})}\BibitemShut {NoStop}%
\bibitem [{\citenamefont {Banerjee}\ \emph {et~al.}(2022)\citenamefont
  {Banerjee}, \citenamefont {Bhattacharyya}, \citenamefont {Drashni},\ and\
  \citenamefont {Pawar}}]{Krylov23}%
  \BibitemOpen
  \bibfield  {author} {\bibinfo {author} {\bibfnamefont {A.}~\bibnamefont
  {Banerjee}}, \bibinfo {author} {\bibfnamefont {A.}~\bibnamefont
  {Bhattacharyya}}, \bibinfo {author} {\bibfnamefont {P.}~\bibnamefont
  {Drashni}}, \ and\ \bibinfo {author} {\bibfnamefont {S.}~\bibnamefont
  {Pawar}},\ }\bibfield  {title} {\bibinfo {title} {From cfts to theories with
  bondi-metzner-sachs symmetries: Complexity and out-of-time-ordered
  correlators},\ }\href {\doibase/10.1103/PhysRevD.106.126022} {\bibfield
  {journal} {\bibinfo  {journal} {Phys. Rev. D}\ }\textbf {\bibinfo {volume}
  {106}},\ \bibinfo {pages} {126022} (\bibinfo {year} {2022})}\BibitemShut
  {NoStop}%
\bibitem [{\citenamefont {Fan}(2022{\natexlab{a}})}]{Krylov24}%
  \BibitemOpen
  \bibfield  {author} {\bibinfo {author} {\bibfnamefont {Z.-Y.}\ \bibnamefont
  {Fan}},\ }\bibfield  {title} {\bibinfo {title} {Universal relation for
  operator complexity},\ }\href {\doibase/10.1103/PhysRevA.105.062210}
  {\bibfield  {journal} {\bibinfo  {journal} {Phys. Rev. A}\ }\textbf {\bibinfo
  {volume} {105}},\ \bibinfo {pages} {062210} (\bibinfo {year}
  {2022}{\natexlab{a}})}\BibitemShut {NoStop}%
\bibitem [{\citenamefont {Fan}(2022{\natexlab{b}})}]{Krylov25}%
  \BibitemOpen
  \bibfield  {author} {\bibinfo {author} {\bibfnamefont {Z.-Y.}\ \bibnamefont
  {Fan}},\ }\bibfield  {title} {\bibinfo {title} {The growth of operator
  entropy in operator growth},\ }\href {\doibase/10.1007/JHEP08(2022)232}
  {\bibfield  {journal} {\bibinfo  {journal} {Journal of High Energy Physics}\
  }\textbf {\bibinfo {volume} {2022}},\ \bibinfo {pages} {232} (\bibinfo {year}
  {2022}{\natexlab{b}})}\BibitemShut {NoStop}%
\bibitem [{\citenamefont {Rabinovici}\ \emph
  {et~al.}(2022{\natexlab{b}})\citenamefont {Rabinovici}, \citenamefont
  {S{\'a}nchez-Garrido}, \citenamefont {Shir},\ and\ \citenamefont
  {Sonner}}]{Krylov26}%
  \BibitemOpen
  \bibfield  {author} {\bibinfo {author} {\bibfnamefont {E.}~\bibnamefont
  {Rabinovici}}, \bibinfo {author} {\bibfnamefont {A.}~\bibnamefont
  {S{\'a}nchez-Garrido}}, \bibinfo {author} {\bibfnamefont {R.}~\bibnamefont
  {Shir}}, \ and\ \bibinfo {author} {\bibfnamefont {J.}~\bibnamefont
  {Sonner}},\ }\bibfield  {title} {\bibinfo {title} {Krylov complexity from
  integrability to chaos},\ }\href {\doibase/10.1007/JHEP07(2022)151}
  {\bibfield  {journal} {\bibinfo  {journal} {Journal of High Energy Physics}\
  }\textbf {\bibinfo {volume} {2022}},\ \bibinfo {pages} {151} (\bibinfo {year}
  {2022}{\natexlab{b}})}\BibitemShut {NoStop}%
\bibitem [{\citenamefont {Bhattacharyya}\ \emph {et~al.}(2023)\citenamefont
  {Bhattacharyya}, \citenamefont {Ghosh},\ and\ \citenamefont
  {Nandi}}]{Hubbard3}%
  \BibitemOpen
  \bibfield  {author} {\bibinfo {author} {\bibfnamefont {A.}~\bibnamefont
  {Bhattacharyya}}, \bibinfo {author} {\bibfnamefont {D.}~\bibnamefont
  {Ghosh}}, \ and\ \bibinfo {author} {\bibfnamefont {P.}~\bibnamefont
  {Nandi}},\ }\bibfield  {title} {\bibinfo {title} {Operator growth and krylov
  complexity in bose-hubbard model},\ }\href {\doibase/10.1007/jhep12(2023)112}
  {\bibfield  {journal} {\bibinfo  {journal} {Journal of High Energy Physics}\
  }\textbf {\bibinfo {volume} {2023}},\ \bibinfo {pages} {112} (\bibinfo {year}
  {2023})}\BibitemShut {NoStop}%
\bibitem [{\citenamefont {Nandy}\ \emph {et~al.}(2024)\citenamefont {Nandy},
  \citenamefont {Mukherjee}, \citenamefont {Bhattacharyya},\ and\ \citenamefont
  {Banerjee}}]{KC_MBS}%
  \BibitemOpen
  \bibfield  {author} {\bibinfo {author} {\bibfnamefont {S.}~\bibnamefont
  {Nandy}}, \bibinfo {author} {\bibfnamefont {B.}~\bibnamefont {Mukherjee}},
  \bibinfo {author} {\bibfnamefont {A.}~\bibnamefont {Bhattacharyya}}, \ and\
  \bibinfo {author} {\bibfnamefont {A.}~\bibnamefont {Banerjee}},\ }\bibfield
  {title} {\bibinfo {title} {Quantum state complexity meets many-body scars},\
  }\href {\doibase/10.1088/1361-648X/ad1a7b} {\bibfield  {journal} {\bibinfo
  {journal} {Journal of Physics: Condensed Matter}\ }\textbf {\bibinfo {volume}
  {36}},\ \bibinfo {pages} {155601} (\bibinfo {year} {2024})}\BibitemShut
  {NoStop}%
\bibitem [{\citenamefont {Jian}\ \emph {et~al.}(2021)\citenamefont {Jian},
  \citenamefont {Swingle},\ and\ \citenamefont {Xian}}]{Krylov8}%
  \BibitemOpen
  \bibfield  {author} {\bibinfo {author} {\bibfnamefont {S.-K.}\ \bibnamefont
  {Jian}}, \bibinfo {author} {\bibfnamefont {B.}~\bibnamefont {Swingle}}, \
  and\ \bibinfo {author} {\bibfnamefont {Z.-Y.}\ \bibnamefont {Xian}},\
  }\bibfield  {title} {\bibinfo {title} {Complexity growth of operators in the
  syk model and in jt gravity},\ }\href {\doibase/10.1007/JHEP03(2021)014}
  {\bibfield  {journal} {\bibinfo  {journal} {Journal of High Energy Physics}\
  }\textbf {\bibinfo {volume} {2021}},\ \bibinfo {pages} {14} (\bibinfo {year}
  {2021})}\BibitemShut {NoStop}%
\bibitem [{\citenamefont {Dymarsky}\ and\ \citenamefont
  {Smolkin}(2021)}]{Krylov10}%
  \BibitemOpen
  \bibfield  {author} {\bibinfo {author} {\bibfnamefont {A.}~\bibnamefont
  {Dymarsky}}\ and\ \bibinfo {author} {\bibfnamefont {M.}~\bibnamefont
  {Smolkin}},\ }\bibfield  {title} {\bibinfo {title} {Krylov complexity in
  conformal field theory},\ }\href {\doibase/10.1103/PhysRevD.104.L081702}
  {\bibfield  {journal} {\bibinfo  {journal} {Phys. Rev. D}\ }\textbf {\bibinfo
  {volume} {104}},\ \bibinfo {pages} {L081702} (\bibinfo {year}
  {2021})}\BibitemShut {NoStop}%
\bibitem [{\citenamefont {Balasubramanian}\ \emph {et~al.}(2022)\citenamefont
  {Balasubramanian}, \citenamefont {Caputa}, \citenamefont {Magan},\ and\
  \citenamefont {Wu}}]{Krylov17}%
  \BibitemOpen
  \bibfield  {author} {\bibinfo {author} {\bibfnamefont {V.}~\bibnamefont
  {Balasubramanian}}, \bibinfo {author} {\bibfnamefont {P.}~\bibnamefont
  {Caputa}}, \bibinfo {author} {\bibfnamefont {J.~M.}\ \bibnamefont {Magan}}, \
  and\ \bibinfo {author} {\bibfnamefont {Q.}~\bibnamefont {Wu}},\ }\bibfield
  {title} {\bibinfo {title} {Quantum chaos and the complexity of spread of
  states},\ }\href {\doibase/10.1103/PhysRevD.106.046007} {\bibfield  {journal}
  {\bibinfo  {journal} {Phys. Rev. D}\ }\textbf {\bibinfo {volume} {106}},\
  \bibinfo {pages} {046007} (\bibinfo {year} {2022})}\BibitemShut {NoStop}%
\bibitem [{\citenamefont {Carabba}\ \emph {et~al.}(2022)\citenamefont
  {Carabba}, \citenamefont {Hörnedal},\ and\ \citenamefont {Campo}}]{autoC3}%
  \BibitemOpen
  \bibfield  {author} {\bibinfo {author} {\bibfnamefont {N.}~\bibnamefont
  {Carabba}}, \bibinfo {author} {\bibfnamefont {N.}~\bibnamefont {Hörnedal}},
  \ and\ \bibinfo {author} {\bibfnamefont {A.~d.}\ \bibnamefont {Campo}},\
  }\bibfield  {title} {\bibinfo {title} {Quantum speed limits on operator flows
  and correlation functions},\ }\href {\doibase/10.22331/q-2022-12-22-884}
  {\bibfield  {journal} {\bibinfo  {journal} {Quantum}\ }\textbf {\bibinfo
  {volume} {6}},\ \bibinfo {pages} {884} (\bibinfo {year} {2022})}\BibitemShut
  {NoStop}%
\bibitem [{\citenamefont {Kubo}(1957)}]{autoC1}%
  \BibitemOpen
  \bibfield  {author} {\bibinfo {author} {\bibfnamefont {R.}~\bibnamefont
  {Kubo}},\ }\bibfield  {title} {\bibinfo {title} {Statistical-mechanical
  theory of irreversible processes. i. general theory and simple applications
  to magnetic and conduction problems},\ }\href {\doibase/10.1143/JPSJ.12.570}
  {\bibfield  {journal} {\bibinfo  {journal} {Journal of the Physical Society
  of Japan}\ }\textbf {\bibinfo {volume} {12}},\ \bibinfo {pages} {570}
  (\bibinfo {year} {1957})}\BibitemShut {NoStop}%
\bibitem [{\citenamefont {Alhambra}\ \emph {et~al.}(2020)\citenamefont
  {Alhambra}, \citenamefont {Riddell},\ and\ \citenamefont
  {Garc\'{\i}a-Pintos}}]{autoC2}%
  \BibitemOpen
  \bibfield  {author} {\bibinfo {author} {\bibfnamefont {A.~M.}\ \bibnamefont
  {Alhambra}}, \bibinfo {author} {\bibfnamefont {J.}~\bibnamefont {Riddell}}, \
  and\ \bibinfo {author} {\bibfnamefont {L.~P.}\ \bibnamefont
  {Garc\'{\i}a-Pintos}},\ }\bibfield  {title} {\bibinfo {title} {Time evolution
  of correlation functions in quantum many-body systems},\ }\href
  {\doibase/10.1103/PhysRevLett.124.110605} {\bibfield  {journal} {\bibinfo
  {journal} {Phys. Rev. Lett.}\ }\textbf {\bibinfo {volume} {124}},\ \bibinfo
  {pages} {110605} (\bibinfo {year} {2020})}\BibitemShut {NoStop}%
\bibitem [{\citenamefont {Zhang}\ and\ \citenamefont {Zhai}(2024)}]{UHAF-z}%
  \BibitemOpen
  \bibfield  {author} {\bibinfo {author} {\bibfnamefont {R.}~\bibnamefont
  {Zhang}}\ and\ \bibinfo {author} {\bibfnamefont {H.}~\bibnamefont {Zhai}},\
  }\bibfield  {title} {\bibinfo {title} {Universal hypothesis of
  autocorrelation function from krylov complexity},\ }\href
  {http://dx.doi.org/10.1007/s44214-024-00054-4} {\bibfield  {journal}
  {\bibinfo  {journal} {Quantum Frontiers}\ }\textbf {\bibinfo {volume} {3}},\
  \bibinfo {pages} {7} (\bibinfo {year} {2024})}\BibitemShut {NoStop}%
\bibitem [{\citenamefont {Caputa}\ \emph {et~al.}(2022)\citenamefont {Caputa},
  \citenamefont {Magan},\ and\ \citenamefont {Patramanis}}]{Geometry_Krylov1}%
  \BibitemOpen
  \bibfield  {author} {\bibinfo {author} {\bibfnamefont {P.}~\bibnamefont
  {Caputa}}, \bibinfo {author} {\bibfnamefont {J.~M.}\ \bibnamefont {Magan}}, \
  and\ \bibinfo {author} {\bibfnamefont {D.}~\bibnamefont {Patramanis}},\
  }\bibfield  {title} {\bibinfo {title} {Geometry of krylov complexity},\
  }\href {\doibase/10.1103/PhysRevResearch.4.013041} {\bibfield  {journal}
  {\bibinfo  {journal} {Phys. Rev. Research}\ }\textbf {\bibinfo {volume}
  {4}},\ \bibinfo {pages} {013041} (\bibinfo {year} {2022})}\BibitemShut
  {NoStop}%
\bibitem [{\citenamefont {Lv}\ \emph {et~al.}(2023)\citenamefont {Lv},
  \citenamefont {Zhang},\ and\ \citenamefont {Zhou}}]{CC-KC}%
  \BibitemOpen
  \bibfield  {author} {\bibinfo {author} {\bibfnamefont {C.}~\bibnamefont
  {Lv}}, \bibinfo {author} {\bibfnamefont {R.}~\bibnamefont {Zhang}}, \ and\
  \bibinfo {author} {\bibfnamefont {Q.}~\bibnamefont {Zhou}},\ }\href@noop {}
  {\bibinfo {title} {Building krylov complexity from circuit complexity},\ }
  (\bibinfo {year} {2023}),\ \Eprint {https://arxiv.org/abs/2303.07343}
  {arXiv:2303.07343 [quant-ph]} \BibitemShut {NoStop}%
\bibitem [{\citenamefont {Hörnedal}\ \emph {et~al.}(2023)\citenamefont
  {Hörnedal}, \citenamefont {Carabba}, \citenamefont {Takahashi},\ and\
  \citenamefont {del Campo}}]{Geometry_Krylov2}%
  \BibitemOpen
  \bibfield  {author} {\bibinfo {author} {\bibfnamefont {N.}~\bibnamefont
  {Hörnedal}}, \bibinfo {author} {\bibfnamefont {N.}~\bibnamefont {Carabba}},
  \bibinfo {author} {\bibfnamefont {K.}~\bibnamefont {Takahashi}}, \ and\
  \bibinfo {author} {\bibfnamefont {A.}~\bibnamefont {del Campo}},\ }\bibfield
  {title} {\bibinfo {title} {Geometric operator quantum speed limit, wegner
  hamiltonian flow and operator growth},\ }\href
  {\doibase/10.22331/q-2023-07-11-1055} {\bibfield  {journal} {\bibinfo
  {journal} {Quantum}\ }\textbf {\bibinfo {volume} {7}},\ \bibinfo {pages}
  {1055} (\bibinfo {year} {2023})}\BibitemShut {NoStop}%
\bibitem [{\citenamefont {Camargo}\ \emph {et~al.}(2023)\citenamefont
  {Camargo}, \citenamefont {Jahnke}, \citenamefont {Kim},\ and\ \citenamefont
  {Nishida}}]{Temperature1}%
  \BibitemOpen
  \bibfield  {author} {\bibinfo {author} {\bibfnamefont {H.~A.}\ \bibnamefont
  {Camargo}}, \bibinfo {author} {\bibfnamefont {V.}~\bibnamefont {Jahnke}},
  \bibinfo {author} {\bibfnamefont {K.-Y.}\ \bibnamefont {Kim}}, \ and\
  \bibinfo {author} {\bibfnamefont {M.}~\bibnamefont {Nishida}},\ }\bibfield
  {title} {\bibinfo {title} {Krylov complexity in free and interacting scalar
  field theories with bounded power spectrum},\ }\href
  {http://dx.doi.org/10.1007/JHEP05(2023)226} {\bibfield  {journal} {\bibinfo
  {journal} {Journal of High Energy Physics}\ }\textbf {\bibinfo {volume}
  {2023}} (\bibinfo {year} {2023})}\BibitemShut {NoStop}%
\bibitem [{\citenamefont {Camargo}\ \emph {et~al.}(2024)\citenamefont
  {Camargo}, \citenamefont {Jahnke}, \citenamefont {Jeong}, \citenamefont
  {Kim},\ and\ \citenamefont {Nishida}}]{Temperature3}%
  \BibitemOpen
  \bibfield  {author} {\bibinfo {author} {\bibfnamefont {H.~A.}\ \bibnamefont
  {Camargo}}, \bibinfo {author} {\bibfnamefont {V.}~\bibnamefont {Jahnke}},
  \bibinfo {author} {\bibfnamefont {H.-S.}\ \bibnamefont {Jeong}}, \bibinfo
  {author} {\bibfnamefont {K.-Y.}\ \bibnamefont {Kim}}, \ and\ \bibinfo
  {author} {\bibfnamefont {M.}~\bibnamefont {Nishida}},\ }\bibfield  {title}
  {\bibinfo {title} {Spectral and krylov complexity in billiard systems},\
  }\href {\doibase/10.1103/PhysRevD.109.046017} {\bibfield  {journal} {\bibinfo
   {journal} {Phys. Rev. D}\ }\textbf {\bibinfo {volume} {109}},\ \bibinfo
  {pages} {046017} (\bibinfo {year} {2024})}\BibitemShut {NoStop}%
\bibitem [{\citenamefont {Tang}(2023)}]{Temperature2}%
  \BibitemOpen
  \bibfield  {author} {\bibinfo {author} {\bibfnamefont {H.}~\bibnamefont
  {Tang}},\ }\href@noop {} {\bibinfo {title} {Operator krylov complexity in
  random matrix theory},\ } (\bibinfo {year} {2023}),\ \Eprint
  {https://arxiv.org/abs/2312.17416} {arXiv:2312.17416 [hep-th]} \BibitemShut
  {NoStop}%
\bibitem [{\citenamefont {Edwards}\ and\ \citenamefont
  {Thouless}(1972)}]{THt1}%
  \BibitemOpen
  \bibfield  {author} {\bibinfo {author} {\bibfnamefont {J.~T.}\ \bibnamefont
  {Edwards}}\ and\ \bibinfo {author} {\bibfnamefont {D.~J.}\ \bibnamefont
  {Thouless}},\ }\bibfield  {title} {\bibinfo {title} {Numerical studies of
  localization in disordered systems},\ }\href
  {\doibase/10.1088/0022-3719/5/8/007} {\bibfield  {journal} {\bibinfo
  {journal} {Journal of Physics C: Solid State Physics}\ }\textbf {\bibinfo
  {volume} {5}},\ \bibinfo {pages} {807} (\bibinfo {year} {1972})}\BibitemShut
  {NoStop}%
\bibitem [{\citenamefont {Thouless}(1974)}]{Tht2}%
  \BibitemOpen
  \bibfield  {author} {\bibinfo {author} {\bibfnamefont {D.}~\bibnamefont
  {Thouless}},\ }\bibfield  {title} {\bibinfo {title} {Electrons in disordered
  systems and the theory of localization},\ }\href
  {\doibase/https://doi.org/10.1016/0370-1573(74)90029-5} {\bibfield  {journal}
  {\bibinfo  {journal} {Physics Reports}\ }\textbf {\bibinfo {volume} {13}},\
  \bibinfo {pages} {93} (\bibinfo {year} {1974})}\BibitemShut {NoStop}%
\bibitem [{\citenamefont {D’Alessio}\ \emph {et~al.}(2016)\citenamefont
  {D’Alessio}, \citenamefont {Kafri}, \citenamefont {Polkovnikov},\ and\
  \citenamefont {Rigol}}]{Tht3}%
  \BibitemOpen
  \bibfield  {author} {\bibinfo {author} {\bibfnamefont {L.}~\bibnamefont
  {D’Alessio}}, \bibinfo {author} {\bibfnamefont {Y.}~\bibnamefont {Kafri}},
  \bibinfo {author} {\bibfnamefont {A.}~\bibnamefont {Polkovnikov}}, \ and\
  \bibinfo {author} {\bibfnamefont {M.}~\bibnamefont {Rigol}},\ }\bibfield
  {title} {\bibinfo {title} {From quantum chaos and eigenstate thermalization
  to statistical mechanics and thermodynamics},\ }\href
  {\doibase/10.1080/00018732.2016.1198134} {\bibfield  {journal} {\bibinfo
  {journal} {Advances in Physics}\ }\textbf {\bibinfo {volume} {65}},\ \bibinfo
  {pages} {239–362} (\bibinfo {year} {2016})}\BibitemShut {NoStop}%
\bibitem [{\citenamefont {Vasilyev}\ \emph {et~al.}(2020)\citenamefont
  {Vasilyev}, \citenamefont {Grankin}, \citenamefont {Baranov}, \citenamefont
  {Sieberer},\ and\ \citenamefont {Zoller}}]{MSFF2020}%
  \BibitemOpen
  \bibfield  {author} {\bibinfo {author} {\bibfnamefont {D.~V.}\ \bibnamefont
  {Vasilyev}}, \bibinfo {author} {\bibfnamefont {A.}~\bibnamefont {Grankin}},
  \bibinfo {author} {\bibfnamefont {M.~A.}\ \bibnamefont {Baranov}}, \bibinfo
  {author} {\bibfnamefont {L.~M.}\ \bibnamefont {Sieberer}}, \ and\ \bibinfo
  {author} {\bibfnamefont {P.}~\bibnamefont {Zoller}},\ }\bibfield  {title}
  {\bibinfo {title} {Monitoring quantum simulators via quantum nondemolition
  couplings to atomic clock qubits},\ }\href
  {\doibase/10.1103/PRXQuantum.1.020302} {\bibfield  {journal} {\bibinfo
  {journal} {PRX Quantum}\ }\textbf {\bibinfo {volume} {1}},\ \bibinfo {pages}
  {020302} (\bibinfo {year} {2020})}\BibitemShut {NoStop}%
\bibitem [{\citenamefont {Noh}(2021)}]{Noh}%
  \BibitemOpen
  \bibfield  {author} {\bibinfo {author} {\bibfnamefont {J.~D.}\ \bibnamefont
  {Noh}},\ }\bibfield  {title} {\bibinfo {title} {Operator growth in the
  transverse-field ising spin chain with integrability-breaking longitudinal
  field},\ }\href {\doibase/10.1103/PhysRevE.104.034112} {\bibfield  {journal}
  {\bibinfo  {journal} {Phys. Rev. E}\ }\textbf {\bibinfo {volume} {104}},\
  \bibinfo {pages} {034112} (\bibinfo {year} {2021})}\BibitemShut {NoStop}%
\bibitem [{\citenamefont {Nivedita}\ \emph {et~al.}(2020)\citenamefont
  {Nivedita}, \citenamefont {Shackleton},\ and\ \citenamefont
  {Sachdev}}]{RIsing1}%
  \BibitemOpen
  \bibfield  {author} {\bibinfo {author} {\bibnamefont {Nivedita}}, \bibinfo
  {author} {\bibfnamefont {H.}~\bibnamefont {Shackleton}}, \ and\ \bibinfo
  {author} {\bibfnamefont {S.}~\bibnamefont {Sachdev}},\ }\bibfield  {title}
  {\bibinfo {title} {Spectral form factors of clean and random quantum ising
  chains},\ }\href {\doibase/10.1103/PhysRevE.101.042136} {\bibfield  {journal}
  {\bibinfo  {journal} {Phys. Rev. E}\ }\textbf {\bibinfo {volume} {101}},\
  \bibinfo {pages} {042136} (\bibinfo {year} {2020})}\BibitemShut {NoStop}%
\bibitem [{\citenamefont {De~Tomasi}\ \emph {et~al.}(2021)\citenamefont
  {De~Tomasi}, \citenamefont {Khaymovich}, \citenamefont {Pollmann},\ and\
  \citenamefont {Warzel}}]{RIsingL}%
  \BibitemOpen
  \bibfield  {author} {\bibinfo {author} {\bibfnamefont {G.}~\bibnamefont
  {De~Tomasi}}, \bibinfo {author} {\bibfnamefont {I.~M.}\ \bibnamefont
  {Khaymovich}}, \bibinfo {author} {\bibfnamefont {F.}~\bibnamefont
  {Pollmann}}, \ and\ \bibinfo {author} {\bibfnamefont {S.}~\bibnamefont
  {Warzel}},\ }\bibfield  {title} {\bibinfo {title} {Rare thermal bubbles at
  the many-body localization transition from the fock space point of view},\
  }\href {\doibase/10.1103/PhysRevB.104.024202} {\bibfield  {journal} {\bibinfo
   {journal} {Phys. Rev. B}\ }\textbf {\bibinfo {volume} {104}},\ \bibinfo
  {pages} {024202} (\bibinfo {year} {2021})}\BibitemShut {NoStop}%
\end{thebibliography}
%

\end{document}